\documentclass[10pt,journal,compsoc]{IEEEtran}
\ifCLASSOPTIONcompsoc
  \usepackage[nocompress]{cite}
\else
  \usepackage{cite}
\fi
\ifCLASSINFOpdf
   \usepackage[pdftex]{graphicx}
\else
   \usepackage[dvips]{graphicx}
\fi
\usepackage{amsmath}
\ifCLASSOPTIONcompsoc
  \usepackage[caption=false,font=footnotesize,labelfont=sf,textfont=sf]{subfig}
\else
  \usepackage[caption=false,font=footnotesize]{subfig}
\fi
\usepackage{url}

\usepackage{xcolor,colortbl}
\usepackage{bm}
\usepackage{booktabs} 
\usepackage{tabularx}
\usepackage{multirow}
\usepackage{comment}

\usepackage{listings}
  \lstdefinelanguage{diff}{
    basicstyle=\ttfamily\footnotesize,
    morecomment=[f][\color{gray}]{@@},
    morecomment=[f][\color{red}]{-\ },
    morecomment=[f][\color{green}]{+\ },
  }

\usepackage{tcolorbox}

\newcommand{\system}{{\textsf{Ratchet}}}
\newcommand{\dataset}{\url{https://github.com/hideakihata/NMTbasedCorrectivePatchGenerationDataset}}

\begin{document}
%
\title{Learning to Generate Corrective Patches Using Neural Machine Translation}
%
%
%
%

\author{Hideaki~Hata,~\IEEEmembership{Member,~IEEE,}
        Emad~Shihab,~\IEEEmembership{Senior Member,~IEEE,}
        and~Graham~Neubig
\IEEEcompsocitemizethanks{
\IEEEcompsocthanksitem H. Hata is with Division of Information Science, Nara Institute of Science and Technology, Nara, 6300192, Japan.\protect\\
E-mail: hata@is.naist.jp
\IEEEcompsocthanksitem E. Shihab is with Department of Computer Science and Software Engineering, Concordia University, Montreal, QC H3G 1M8, Canada.
E-mail: eshihab@cse.concordia.ca
\IEEEcompsocthanksitem G. Neubig is with Language Technology Institute, Carnegie Mellon University, Pittsburgh, PA 15213-3891, USA.
E-mail: gneubig@cs.cmu.edu
}
\thanks{Manuscript received April 19, 2005; revised August 26, 2015.}}

%
%

\markboth{Journal of \LaTeX\ Class Files,~Vol.~14, No.~8, August~2015}%
{Shell \MakeLowercase{\textit{et al.}}: Bare Demo of IEEEtran.cls for Computer Society Journals}
%



\IEEEtitleabstractindextext{%
\begin{abstract}
Bug fixing is generally a manually-intensive task. However, recent work has proposed the idea of automated program repair, which aims to repair (at least a subset of) bugs in different ways such as code mutation, etc. Following in the same line of work as automated bug repair, in this paper we aim to leverage past fixes to propose fixes of current/future bugs. Specifically, we propose \system, a corrective patch generation system using neural machine translation. By learning corresponding pre-correction and post-correction code in past fixes with a neural sequence-to-sequence model, \system~is able to generate a fix code for a given bug-prone code query. We perform an empirical study with five open source projects, namely Ambari, Camel, Hadoop, Jetty and Wicket, to evaluate the effectiveness of \system. Our findings show that \system~can generate syntactically valid statements 98.7\% of the time, and achieve an F1-measure between 0.29 -- 0.83 with respect to the actual fixes adopted in the code base. 
In addition, we perform a qualitative validation using 20 participants to see whether the generated statements can be helpful in correcting bugs. Our survey showed that \system's output was considered to be helpful in fixing the bugs on many occasions, even if fix was not 100\% correct.
\end{abstract}

\begin{IEEEkeywords}
patch generation, corrective patches, neural machine translation, change reuse.
\end{IEEEkeywords}}

\maketitle

\IEEEdisplaynontitleabstractindextext

%
\IEEEpeerreviewmaketitle

\IEEEraisesectionheading{\section{Introduction}\label{sec:introduction}}

%
%
%
%

 

Most software bug fixing tasks are manual and tedious. Recently, a number of techniques related to automated program repair have been proposed to help automate and reduce the burden of some of these tasks~\cite{Pei:2014:AFP:2693206.2693287,
LeGoues:2012:GGM:2122269.2122544,
LeGoues:2012:SSA:2337223.2337225,
Liu:2012:AAF:2337223.2337259,
Nguyen:2013:SPR:2486788.2486890,
Coker:2013:PTF:2486788.2486892,
Weimer:2013:LPE:3107656.3107702,
Qi:2014:SRS:2568225.2568254,
Kaleeswaran:2014:MAS:2568225.2568258,
OcarizaJr.:2014:VSF:2568225.2568257,
Liu:2014:GCF:2635868.2635881,
Lin:2014:ARM:2610384.2610398,
Mechtaev:2015:DLS:2818754.2818811,
Gao:2015:SMF:2818754.2818812,
Tan:2015:RAR:2818754.2818813,
Long:2015:SPR:2786805.2786811,
Mechtaev:2016:ASM:2884781.2884807,
Xuan:2017:NAR:3071893.3071964,
Xiong:2017:PCS:3097368.3097418,
Le:2017:SSS:3106237.3106309,
Saha:2017:EEO:3155562.3155643,
Hua:2018:TPP:3180155.3180245,
Mechtaev:2018:SPR:3180155.3180247,
Monperrus:2018:ASR:3177787.3105906,
8089448}.
These systems are also seeing practical use.
For example, Facebook has announced that they started applying a system of automated program repair called SapFix in their large-scale products~\cite{safix}.



However, there are limitations in current approaches to automated program repair. First, there is a risk of overfitting to the training set (and breaking under tested functionality) in patch generation, especially generated tests tends to lead overfitting compared to human-generated, requirements-based test suites~\cite{Smith:2015:CWD:2786805.2786825}.
Second, correct patches may not exist in the search space, or correct patches cannot be generated because the search space is huge~\cite{Long:2016:ASS:2884781.2884872,Wen:2018:CPG:3180155.3180233}.
Several studies address this search space issue by making use of existing human-written patches~\cite{Kim:2013:APG:2486788.2486893,
Martinez:2015:MSR:2727049.2727080,
Ke:2015:RPS:2916135.2916260,
7476644,
Campos:2017:CBP:3200492.3200554,
Long:2017:AIC:3106237.3106253,
Xin:2017:LSC:3155562.3155644,
Jiang:2018:SPR:3213846.3213871}, but those generated patches need to be validated with test suites.
Therefore, investigating techniques that assist in the generation of patches without the need for tests, etc. are needed.
Instead of exploring fix ingredients in the search space (search-based), 
we study the possibility of learning fix ingredients from past fixes (learning-based).


Recently, Neural Machine Translation (NMT) has been proposed and showed promising results in various areas including not only translation between natural languages (such as English and Japanese), but also other NLP tasks such as speech recognition~\cite{chan2016listen}, natural language parsing~\cite{vinyals2015grammar}, and text summarization~\cite{rush2015neuralattention}.
Similar techniques have been applied to code-related tasks~\cite{iyer16summarizing,
ling-EtAl:2016:P16-1,
yin-neubig:2017:Long,
rabinovich-stern-klein:2017:Long,
Gu:2016:DAL:2950290.2950334,
bhatia2016automated,AAAI1714603}.
The notable success of NMT in such a wide variety of tasks can be attributed to several traits:
(1) It is an end-to-end machine learning framework that can be learned effectively from large data -- if we have a large enough data source it is able to learn even complicated tasks such as translation in an effective way;
(2) Unlike previous models for translation such as phrase-based translation~\cite{koehn10smt} (which has also been used in code-related tasks such as language porting~\cite{nguyen2013lexical} and pseudo-code generation~\cite{Oda:2015:LGP:2916135.2916173}), NMT is able to take a holistic look at the entire input and make global decisions about which words or tokens to output.
In particular, for bug fixing we posit this holistic view of the entire hunk of code we attempt to fix is important, and thus focus on approaches using NMT in this work.

Hence, in this paper, we propose \system, a NMT-based technique that generates bug fixes based on prior bug-and-fix examples. To evaluate the effectiveness of the technique, we perform an empirical study with five large software projects, namely Ambari, Camel, Hadoop, Jetty and Wicket. We use the pattern-based patch suggestion inspired by the Plastic Surgery work~\cite{Barr:2014:PSH:2635868.2635898}, as a comparison baseline and examine the effectiveness of our NMT-based technique. In particular, we quantify the number of cases where our NMT-based technique is able to generate a valid fix and how accurate the generated fixes are. Our findings showed that \system~is able to generate a valid statements in 98.7\% of the cases and achieves an F1 measure between 0.29 -- 0.83 with respect to the actual fixes adopted in the code base.
For all five projects, \system~was able to either outperform or perform as well as the baseline.

In addition to the quantitative validation, we also performed a survey with 20 participants to see whether the generated statements can help in correcting a bug (even if they were not 100\% identical to the fix). Our findings through a participant survey show that the fixes generated by \system~are very helpful, even if they were not fully correct (although the correct fixes were most helpful).

There are several recent studies on techniques to generate patches without test cases, which differ from our approach:
\emph{inductive programming} for program synthesis making used of historical change patterns~\cite{Rolim:2017:LSP:3097368.3097417},
additive program transformations using \emph{separation logic} to identify and repair the lack of certain operations on heap issues~\cite{vanTonder:2018:SAP:3180155.3180250}, and
learning fix patterns of \textsf{FindBugs} violations using \emph{convolutional neural networks}~\cite{8565907}.
Similar to our approach, these proposals have learning aspects to generate patches without test cases. Major differences are specific targets (heap properties~\cite{vanTonder:2018:SAP:3180155.3180250} and static analysis tool violations~\cite{8565907}) and/or specific patterns to be learned (specified constraints~\cite{Rolim:2017:LSP:3097368.3097417} and manual fix specifications~\cite{vanTonder:2018:SAP:3180155.3180250}), while \system~learns any statement-level changes in a general NMT framework. Although limiting to specific targets and patterns could be effective for the targeted domains, our approach is able to target daily bug fixing activities.

Our approach can be thought of as a method for ``learning-based automated code changes'' instead of one of automated program repair per se.
Although the setting on automated program repair is expensive, especially for validation~\cite{Saha:2017:EEO:3155562.3155643}, our NMT approach can work lightly for usual repetitive maintenance activities.
As automated program repair research is recommended to focus on difficult and important bugs~\cite{Motwani2018}, research on learning-based automated code changes could support repetitive and similar bug fixing tasks by learning common corrective maintenance activities.
We expect that our approach can be integrated in daily maintenance activities. \system~can recommend generated patches to local code before committing to repositories and to submitted code for reviewing. While it could work in human-involved maintenance processes, we consider our approach is not necessarily an end-to-end bug fixing solution by assessing the correctness of generated patches.

The rest of the paper is organized as follows. Section~\ref{sec:term} presents relevant terminology. Section~\ref{sec:background} provides background about NMT. Section~\ref{sec:approach} details our approach. Section~\ref{sec:exp_setup} sets up our experiments, discussing their design and the data used. Section~\ref{sec:eval} presents our results and Section~\ref{sec:discussion} discusses the generality and some challenges facing NMT-based solutions. Related work is presented and contrasted in Section~\ref{sec:related_work} and Section~\ref{sec:conclusion} concludes the paper.



\section{Terminology}
\label{sec:term}

We use the term, \textit{change hunk}, similar to the previous study by Ray et al.~\cite{Ray:2015:UCC:2820518.2820526}. \textit{A change hunk is a list of
program statements deleted and added contiguously.} In a single commit to a code repository, typically there are multiple change regions
in multiple files. Even in a single file, there can be multiple change regions.
Those changed regions can be identified with \textit{diff}. Although the previous study by Ray et al. included unchanged statements in a change hunk~\cite{Ray:2015:UCC:2820518.2820526}, we do not include them.
We call deleted and added statements \textit{pre-correction} and \textit{post-correction} statements respectively.
In Listing~\ref{lst:hunk}, the red statement is a pre-correction statement and the green statement is a corresponding post-correction statement, and these associated two statements are considered to be a change hunk.

\begin{lstlisting}[caption={An example of a change hunk in bug fixing.},label={lst:hunk},language=diff]
Commit: 44074f6ae03031ab046b1886790fc31e66e2d74e
Author: Willem Ning Jiang
Date: Sat Jun 9 09:24:15 2012 +0000
Message: CAMEL-5348 fix the issue of Uptime

  uptime /= 24;
  long days = (long) uptime;
- long hours = (long) ((uptime - days) * 60);
+ long hours = (long) ((uptime - days) * 24);
  String s = fmtI.format(days)
                    + (days > 1 ? " days" : " day");
\end{lstlisting}

In this study, we are interested in learning transforming patterns between corresponding pre-correction and post-correction statements.
Thus, we ignore change hunks that only contain deleted or added statements. All change hunks studied in this paper are pairs of pre-correction and post-correction statements.

\section{Background}
\label{sec:background}

Neural machine translation (NMT), also called neural sequence-to-sequence models  (seq2seq)~\cite{kalchbrenner13rnntm,sutskever14sequencetosequence,neubig2017neural} is a method for converting one input sequence $\bm{x}$ into another output sequence $\bm{y}$ using neural networks. As the name suggests, the method was first conceived for and tested on machine translation; for converting one natural language (e.g. English) into another (e.g. French). However, because these methods can work on essentially any problem of converting one sequence into another, they have also been applied to a wide variety of other tasks such as speech recognition~\cite{chan2016listen}, natural language parsing~\cite{vinyals2015grammar}, and text summarization~\cite{rush2015neuralattention}. They have also seen applications to software for generation of natural language comments from code~\cite{iyer16summarizing}, generation of code from natural language~\cite{ling-EtAl:2016:P16-1,yin-neubig:2017:Long,rabinovich-stern-klein:2017:Long}, generation of API sequences~\cite{Gu:2016:DAL:2950290.2950334}, and suggesting fixes to learner code in programming MOOCs~\cite{bhatia2016automated,AAAI1714603}.

In this section we briefly overview neural networks, then explain NMT in detail.

\subsection{Neural Networks}

Neural networks~\cite{goodfellow2016deep}, put simply, are a complicated function that is composed of simpler component parts that each have parameters that control their behavior.
One common example of such a function is the simple multi-layer calculation below, which converts an input vector $\bm{x}$ into an output vector $\bm{y}$:
\begin{align}
\bm{h} & = tanh(W_1 \bm{x} + b_{1}) \nonumber \\
\bm{y} & = W_2 \bm{h} + b_{2}. \label{eq:nn}
\end{align}
Here, $W_1$ and $W_2$ are parameter matrices, and $b_1$ and $b_2$ are parameter vectors (called \textit{bias} vectors).
Importantly, the vector $\bm{h}$ is a \textit{hidden} layer of the neural network, which results from multiplying $W_1$, adding $b_1$, then taking the hyperbolic tangent with respect to the input.
This hidden layer plays an essential role in neural networks, as it allows the network to automatically discover features of the input that may be useful in predicting $\bm{y}$.%
\footnote{In fact, by adding this hidden layer, a simple function such as the above is able to accurately perform any prediction task given a large enough $\bm{h}$ and enough training data, and thus neural networks are known as \textit{universal function approximators}~\cite{hornik1989multilayer}.}

Because neural networks have parameters ($W_1$, $b_1$, etc.) that specify their behavior, it is necessary to learn these parameters from training data.
In general, we do so by calculating how well we do in predicting the correct answer $\bm{y}'$ provided by the training data, and modify the parameters to increase our prediction accuracy.
Formally, we do so by calculating a \textit{loss function} $\ell(\bm{y}, \bm{y'})$ which will (generally) be 0 if we predict perfectly, and higher if we're not doing a good job at prediction.
We then take the derivative of this loss function with respect to the parameters, e.g. $\frac{\partial \ell(\bm{y}, \bm{y'})}{\partial W_1}$, and move the parameters in the direction to reduce the loss function, e.g.
\begin{equation}
W_1 \leftarrow W_1 - \alpha \frac{\partial \ell(\bm{y}, \bm{y'})}{\partial W_1},
\end{equation}
where $\alpha$ is a learning rate that controls how big of a step we take after every update.

The main difficulty here is that we must calculate derivatives $\frac{\partial \ell(\bm{y}, \bm{y'})}{\partial W_1}$.
Even for a relatively simple function such as the one in (\ref{eq:nn}), calculating the derivative by hand can be cumbersome.
Fortunately, this problem can be solved through a process of \textit{back-propagation} (or \textit{auto-differentiation}), which calculates the derivative of the whole complicated function by successively calculating derivatives of the smaller functions and multiplying them together using the chain rule~\cite{rumelhart1988learning}.
Thus, it becomes possible to train arbitrarily complicated functions, as long as they are composed of simple component parts that can be differentiated, and a number of software libraries such as TensorFlow~\cite{abadi2016tensorflow} and DyNet~\cite{neubig17dynet} make it possible to easily do so within applications.

\subsection{Neural Machine Translation}

NMT is an example of applying a complicated function learnable by neural nets and using it to solve a complicated problem: translation.
To generate an output $\bm{y}$ (e.g. corrected hunk of code) given an input $\bm{x}$, these models incrementally generate each token in the output $y_1, y_2, \ldots, y_{|\bm{y}|}$ one at a time.
For example, if our output is ``\texttt{return this . index}'', the model would first predict and generate ``\texttt{return}'', then ``\texttt{this}'', then ``\texttt{.}'', etc. 
This is done in a probabilistic way by calculating the probability of the first token of the output given the input $P(y_1 \mid \bm{x})$, outputting the token in the vocabulary that maximizes this probability, then calculating the probability of the second token given the first token and the snippet $P(y_2 \mid \bm{x}, y_1)$ and similarly outputting the word with the highest probability, etc.
When training the model, we already know a particular output $\bm{y}$ and want to calculate its probability given a particular snippet $\bm{x}$ so we can update the parameters based on the derivatives of this probability.
To do so, we simply multiply these probabilities together using the chain rule as follows:
\begin{equation}
\label{eq:probabilitycalc}
P(\bm{y} \mid \bm{x}) = P(y_1 \mid \bm{x}) P(y_2 \mid \bm{x}, y_1) P(y_3 \mid \bm{x}, y_1, y_2) \ldots 
\end{equation}

\begin{figure}[!t]
\centering
\includegraphics[width=2.5in]{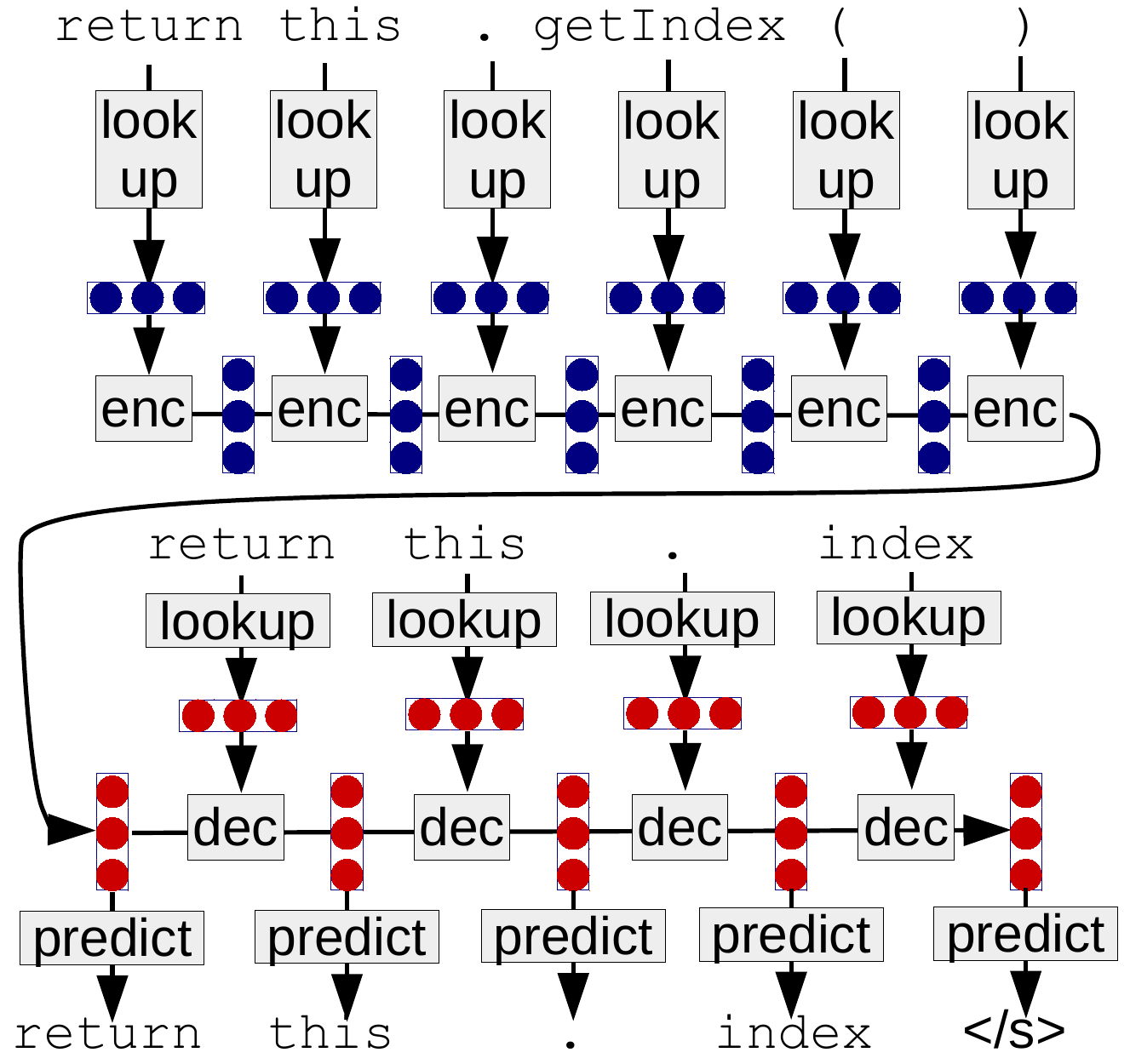}
\caption{An example of NMT encoder-decoder framework used in translation.}
\label{fig:encdec}
\end{figure}

So how do NMT models calculate this probability?
We will explain a basic outline of a basic model called the \textit{encoder-decoder model}~\cite{sutskever14sequencetosequence}, and refer readers to references for details~\cite{sutskever14sequencetosequence,bahdanau15alignandtranslate,neubig2017neural}.
The encoder-decoder model, as shown in Figure \ref{fig:encdec} works in two stages: first it \textit{encodes} the input (in this case $\bm{x}$) into a \textit{hidden vector} of continuous numbers $\bm{h}_x$ using an encoding function
\begin{equation}
\bm{h}_{x,|\bm{x}|} = \text{encode}(\bm{x}).
\end{equation}
This function generally works in two steps: looking up a vector of numbers representing each token (often called ``word embeddings'' or ``word vectors''), then incrementally adding information about these embeddings one token at a time using a particular variety of network called a \textit{recurrent neural network} (RNN).
To take the specific example shown in the figure, at the first time step, we would look up an embedding vector for the first token ``\texttt{return}'', $\bm{e}_1 = \bm{e}_{\texttt{return}}$ and then perform a calculation such as the one below to calculate the hidden vector for the first time step:
\begin{equation}
\bm{h}_{x,1} = \text{tanh}(W_{\text{enc,e}} \bm{e}_1 + b_{\text{enc}}),
\end{equation}
where $W_{\text{enc,e}}$ and $b_{\text{enc}}$ are a matrix and vector that are parameters of the model, and $tanh(\cdot)$ is the hyperbolic tangent function used to ``squish'' the values to be between -1 and 1.%
\footnote{This represents a simple recurrent neural network, but in our actual model we use a more sophisticated version of encoding function called ``long short-term memory'' (LSTM), which performs better on long sequences~\cite{hochreiter97lstm}.}
In the next time step, we would do the same for the symbol ``.'', using its embedding $\bm{e}_2 = \bm{e}_{\texttt{.}}$, and in the calculation from the second step onward we also use the result of the previous calculation (in this case $\bm{h}_{x,2}$):
\begin{equation}
\bm{h}_{x,2} = \text{tanh}(W_{\text{enc,h}} \bm{h}_{x,1} + W_{\text{enc,e}} \bm{e}_2 + b_{\text{enc}}).
\label{eq:rnnwithhidden}
\end{equation}
By using the hidden vector from the previous time step, the RNN is able to ``remember'' features of the previously occurring tokens within this vector, and by repeating this process until the end of the input sequence, it (theoretically) has the ability to remember the entire content of the input within this vector.

Once we have encoded the entire source input, we can then use this encoded vector to predict the first token of the output.
This is generally done by defining the first hidden vector for the output $\bm{h}_{y,0}$ to be equal to the final vector of the input $\bm{h}_{x,|\bm{x}|}$, then multiplying it with another weight vector used for prediction to calculate a score $\bm{g}$ for each token in the output vocabulary:
\begin{equation}
\bm{g}_1 = W_{\text{pred}} \bm{h}_{y,0} + \bm{b}_{\text{pred}}.
\end{equation}
We then predict the actual probability of the first token in the output statement, for example ``\texttt{return}'', by using the \textit{softmax} function, which exponentiates all of the scores in the output vocabulary and then normalizes these scores so that they add to one:
\begin{equation}
P(y_1 = ``\texttt{return}'') = \frac{\text{exp}(g_\text{return})}{\sum_{\tilde{g}} \text{exp}(\tilde{g})}.
\end{equation}
We then calculate a new hidden vector given this input:
\begin{equation}
\bm{h}_{y,1} \text{encode}(y_1 = ``\texttt{return}'', \bm{h}_{y,0}).
\end{equation}
We continue this process recursively until we output a special ``end of hunk'' symbol ``$\langle$/s$\rangle$''.

\textbf{Why NMT models?:}
As mentioned briefly in the intro, NMT models are well-suited to the task of automatic patch generation for a number of reasons.
First, they are an end-to-end probabilistic model that can be trained from parallel datasets of pre- and post-correction code without extra human intervention, making them easy to apply to new datasets or software projects.
Second, they are powerful models that can learn correspondences on a variety of levels; from simple phenomena such as direct token-by-token matches, to soft paraphrases~\cite{socher11dynamic}, to weak correspondences between keywords and large documents for information retrieval~\cite{huang2013learning}. 
Finally, they have demonstrated success in a number of code related tasks as iterated at the beginning of this section, which indicates that they should be useful as part of bug fixing algorithm as well.

\textbf{Attention:}
In addition, we use a NMT model with this basic architecture, with the addition of a feature called \emph{attention}, which, put simply, allows the model to ``focus'' on particular tokens in the input $\bm{x}$ when generating the output $\bm{y}$~\cite{bahdanau15alignandtranslate,luong15effectiveattentional}.
Mathematically, this corresponds to calculating an ``attention vector'' $\bm{a}_j$, given the input hidden vectors $\bm{h}_x$ and the current output hidden vector $\bm{h}_{y,j}$.
This vector consists of values between zero and one, one value for each word in the input, with values closer to one indicating that the model is choosing to focus more on that particular word.
Finally, these values are used to calculate a ``context vector''
\begin{equation}
\bm{c}_j = \sum_{i=1}^{|\bm{x}|} \alpha_{j,i} \bm{h}_{x,i},
\end{equation}
which is used as additional information when calculating score $\bm{g}_j$.
Attention is particularly useful when there are many token-to-token correspondences between the input and output, which we expect to be the case for our patch generation task, where the input and output code are likely to be very similar.
This attention model can be further augmented to allow for exact copies of tokens~\cite{gu-EtAl:2016:P16-1}, or be used to incorporate a dictionary of common token-to-token correspondences (copies or replacements)~\cite{arthur16emnlp}.
In our model, we use the latter, which allows us to both capture the fact that tokens are frequently copied between pre- and post-correction code, and also the fact that some replacements will be particularly common (e.g. \texttt{loadBalancerType} to \texttt{setLoadBalancerType}).
This dictionary is automatically inferred from our training data by running the \texttt{fast\_align} toolkit\footnote{\url{https://github.com/clab/fast_align}}, which can automatically learn such a dictionary from parallel data using probabilistic models \cite{dyer2013simple}.  

\textbf{Implementation details:}
As a specific implementation of the NMT techniques listed above, we use the \textsf{lamtram} toolkit
\cite{neubig15lamtram}.
For reproducibility, we briefly list the parameters below, and interested readers can refer to the references for detail.
As our model we use an encoder-decoder model with multi-layer perceptron attention~\cite{bahdanau15alignandtranslate} and input feeding~\cite{luong15effectiveattentional}, with encoders and decoders using a single layer of 512 LSTM cells~\cite{hochreiter97lstm}.
We use the Adam optimizer~\cite{kingma2014adam} with a learning rate of 0.001 and minibatch size of 2048 words, and decay the learning rate every time the development loss increases.
To prevent overfitting, we use a dropout rate of 0.5~\cite{srivastava2014dropout}.
To generate our outputs, we perform beam search with a beam size of 10.

\begin{figure*}[!t]
\centering
\includegraphics[width=\linewidth]{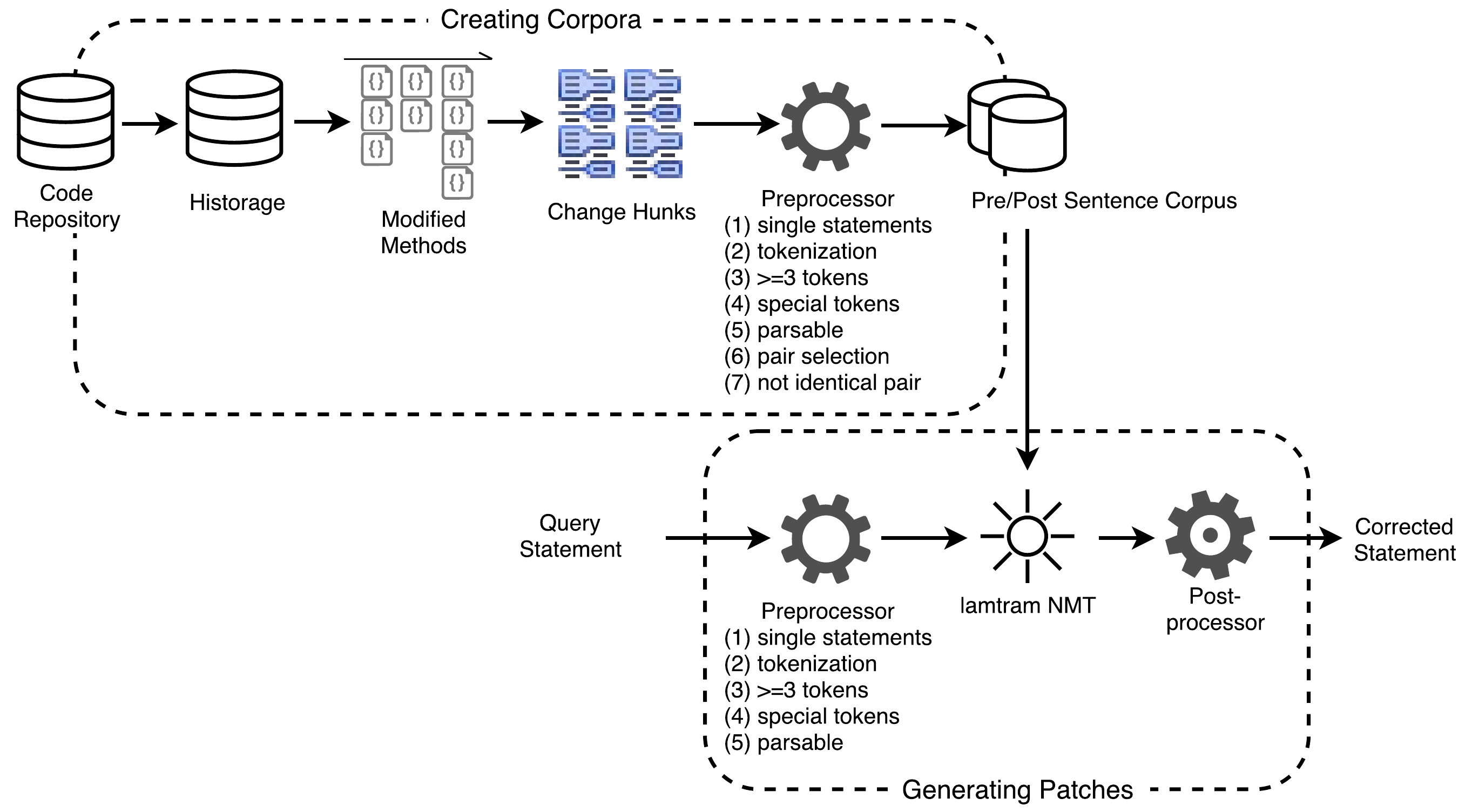}
\caption{Overview of \system, an NMT-based corrective patch generation system.}
\label{fig:overview}
\end{figure*}


\section{Approach}
\label{sec:approach}

The idea of corrective patch generation using NMT considers code changes as translation from pre-correction code to post-correction code.
%
Figure~\ref{fig:overview} provides an overview of our system, \system, which consists of two main parts: creating the training corpora, and generating patches using the trained model. In this paper, we target Java source code and focus on changes within Java methods. Particularly, the granularity of code we target is a statement 
similar to the previous study~\cite{Barr:2014:PSH:2635868.2635898}.
Main focus in this study is preparing appropriate data for a NMT model to learn. To this aim, we build a system to collect fine-grained code change and try ignoring noisy data.



\subsection{Extracting Change Hunks from Code Repositories}
\label{ssec:mining}
In order to create our training corpora, we start by extracting pre- and post-correction statements using a sequence of steps. We detail each of these steps in the following text:


\noindent\textbf{Preparing Historage for method-level histories.}
Since the software repositories store the code modifications at the commit level, our first step is to transform these commits into method-level modifications. To do so, we convert the existing code repositories to  \textsf{historage} repositories~\cite{Hata:2011:HFV:2024445.2024463}. \textsf{Historage} creates a new repository that stores all methods in the logs of the original repository as individual Git objects. In essence, \textsf{historage} is a Git repository that allows us to operate any Git commands as usual.\footnote{We used a tool, called \textsf{kenja}~\cite{Fujiwara:2014:KHS:2597073.2597125} (\url{https://github.com/niyaton/kenja}) to  prepare the \textsf{historage} repositories. Converted \textsf{historage} repositories are now hosted in \textsf{Codosseum} Web service: \url{http://codosseum.naist.jp/}, which is previously presented as \textsf{Kataribe}~\cite{Fujiwara:2014:KHS:2597073.2597125,7550864}.} 

\noindent\textbf{Collecting the modified methods.}
We use the command \texttt{git log --diff-filter=M} on the \textsf{historage} repositories to collect all modified methods in the entire history. The option \texttt{--diff-filter=M} will provide only modified (M) files, which are methods in \textsf{historage} repositories. Since we are interested in training our model on pre- and post-correction statements, we only consider methods that modify code, i.e., not methods that are newly created or completely deleted.

\noindent\textbf{Identifying change hunks.}
As stated in Section~\ref{sec:term}, a change hunk is a pair of pre-correction and post-correction statements. We identify these change hunks from the outputs of the \texttt{git diff}. Since we assume pre-correction statements have been corrected to post-correction statements, we need to identify the corresponding line pairs appropriately.\footnote{The options \texttt{--histogram}, \texttt{--unified=0}, \texttt{--ignore-space- change}, \texttt{--ignore-all-space}, and \texttt{--ignore-blank-lines} are used to apply an advanced diff algorithm, ignore unchanged statements, and ignore trivial changes. An empirical study reported that the histogram algorithm is better than the default diff algorithm in Git~\cite{2019arXiv190202467S}.} 

%
%

\subsection{Preprocessing the Statement Corpora}
\label{ssec:pre}
Before storing the statement pairs as pre-correction and post-correction statement corpora, we perform the following preprocessing steps. As seen in Figure~\ref{fig:overview}, the same processes will be applied to query statements except for the step (6) and (7), which are needed only for creating the corpora. 

	\textbf{(1) Limit to single-statement changes and single-statement queries.}
In this study, we only consider single-statement (one-line) changes. We do so for the following three reasons.
First, previous studies showed that most reusable code is found at the single-statement level~\cite{Nguyen:2013:SRC:3107656.3107682,Barr:2014:PSH:2635868.2635898}. Second, it is difficult to treat multiple statement changes (one-to-many, many-to-one, and many-to-many statement changes) for identifying pairs. Those multiple statement changes can have inappropriate corresponding statements. For example, if there exists one pre-correction statement and two post-correction statements in one change hunk, this change can be a single-statement change and one independent statement insertion. If we consider these statements one pair, the independently inserted statement can be noise in the training data. 
Third, it is difficult to manage past histories associated with multiple statements. Using the command \texttt{git blame} on \textsf{historage}, we identify commits on which deleted lines initially appeared. In general, multiple statements can have different past histories, which makes it difficult to treat those multiple statements as one unit.
For all statement pairs, we collect past history information including the original commit, changed year and deleted year, to be used for our experiments.
Although we apply this filtering, we found that single-statement changes are the majority in our change hunks (as we show later in Figure~\ref{fig:statement_distribution} and Table~\ref{tab:stat}). 


\textbf{(2) Tokenize statements.}
Since the NMT model requires separate tokens as input, we use the \textsf{StreamTokenizer} to tokenize the Java statements.


\textbf{(3) Remove statement pairs or statement queries with less than three tokens.}
We remove statements that have very few tokens (i.e., less than 3) since they are less meaningful. Our observations indication that most such lines only contain opening or closing parenthesis.

\textbf{(4) Replace the contents of method arguments with a special token.}
From our many trials, we realized that a wide variety of the contents of method arguments make it difficult to generate corresponding contents. This is because sometimes method argument contents include tokens that rarely appear.
We replace method and array arguments with a special token, \texttt{arg} and \texttt{val}, respectively.

\textbf{(5) Filter unparseable statement pairs and queries.}
There exist incomplete statements in our collected statements, e.g., when there is a long statement that is written across two lines, and only one line is changed. To remove these incomplete Java statements, we put each statement in a dummy method of a dummy class, and try parsing the class to get an AST using JavaParser.\footnote{JavaParser: \url{http://javaparser.org/}} If we fail to parse classes with either pre- or post-correction statements, we filter out the failed statement pairs 

\textbf{(6) Select post-correction statements from multiple candidates.}
This step is performed to address \textit{the nature of sequential order} in documents.
After collecting all pre- and post-correction statements from the entire history of a code repository, we can have statement pairs that have the same pre-correction statements but different post-correction statements. In order to allow the NMT models to effectively extract relationships or patterns, we chose only one post-correction statement for one pre-correction statement, and remove all other post-correction statements. The idea behind this selection is that it is better to learn from recently and frequently appearing statements. Given a pre-correction statement, we obtain post-correction statements that appeared in the most recent year. Then, from those newer statements, we select statements that most frequently appeared in the entire history. If we cannot break ties, we select the first statement in alphabetical order to make the process deterministic.

\textbf{(7) Remove identical pre- and post-correction statements.}
After the above processes, there exist pairs of identical pre- and post-correction statements. For example, statement pairs from changes only within method arguments, and white space changes. We remove those statement pairs.

\subsection{Post-Processing}
\label{ssec:post}


Since we replace the contents of method arguments and replace it with a special token, the NMT model does not generate method arguments.
However we expect that the method arguments of a query statement can be reused in the generated statement.
Therefore we prepare the following heuristics for new method arguments.
\begin{itemize}
\item Methods that have the same name will have the same method arguments.
\item For chained method calls, arguments are assigned in the same order.
\item If no method argument content is left in a query statement, leave the remaining method call arguments empty.
\end{itemize}

The \textsf{lamtram} toolkit provides scores associated with generated statements with the logarithm of a posteriori probability of output E given input F as $log P(E | F)$. Those scores can be considered as confidences of the results. We empirically determine thresholds and ignore the generated statements with low scores.
In addition, we can also ignore invalid generated statements that cannot be parsed.

\section{Experimental Setup}
\label{sec:exp_setup}
In this section, we discuss our dataset and the design of our experiment.
Particularly, we are interested in examining the viability of our approach in generating bug-fixing statements. 
To do so, we need to identify bug-fixing statement pairs. We discuss the tool used to identify the bug-inducing and bug-fixing commits that are used to determine our bug-fix statement pairs. Then, we provide descriptive statistics about the studied datasets.

\begin{table}[!t]
\renewcommand{\arraystretch}{1.3}
\caption{Descriptive statistics of the studied systems. The number of Java files and methods are from the latest snapshots.}
\label{tab:subject}
\centering
\begin{tabular}{lcrrr}
\toprule
& & \# of & \# of & \# of \\
Project & Period & Commits & Files & Methods \\
\midrule
ambari & Aug-11 to Apr-17 & 14,042 & 2,719 & 29,212 \\
camel & Mar-07 to Jun-17 & 28,668 & 16,889 & 92,839 \\
hadoop & May-09 to Oct-14 & 8,323 & 5,696 & 21,292 \\
jetty & Mar-09 to Apr-16 & 14,167 & 2,668 & 21,172 \\
wicket & Sep-04 to Jun-17 & 19,960 & 5,039 & 16,049 \\
\bottomrule
\end{tabular}
\end{table}

\subsection{Subject Projects}
To perform our case study, we study five projects, namely Apache Ambari, Apache Camel, Apache Hadoop, Eclipse Jetty and Apache Wicket. We chose to study these five projects since they have long development histories and are large projects that contain many commits. Table~\ref{tab:subject} shows the period considered, the number of commits, files and methods in our dataset.

\begin{figure}[!t]
\centering
\includegraphics[width=\linewidth]{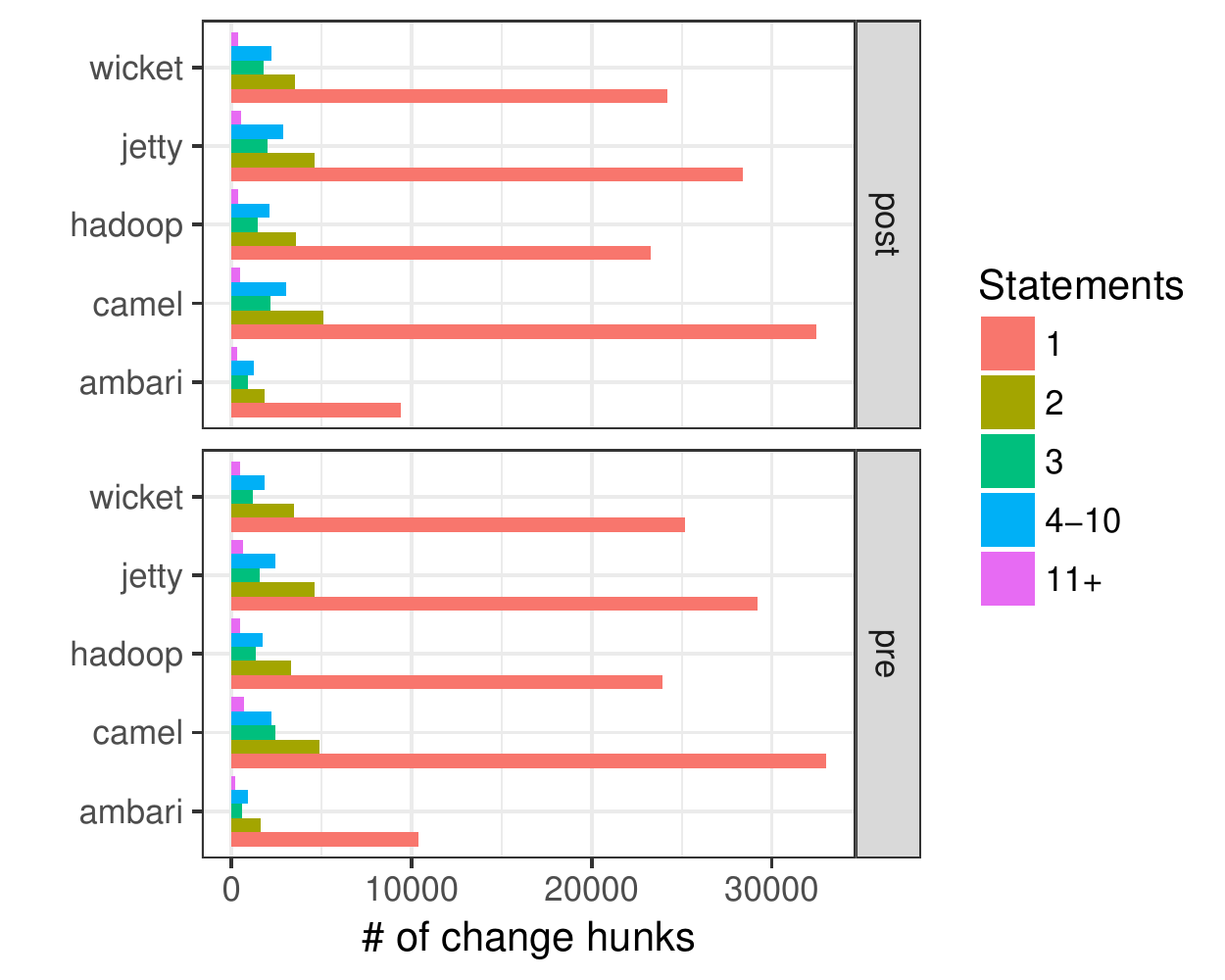}
\caption{The number of change hunks with different numbers of pre- and post-correction statements in the entire periods.}
\label{fig:statement_distribution}
\end{figure}

\begin{table}[!t]
\renewcommand{\arraystretch}{1.3}
\caption{Statistics of change hunks in the entire period.}
\label{tab:stat}
\centering
\begin{tabular}{lrrr}
\toprule
& \# of & \multicolumn{2}{r}{\#, (\%) of Hunks} \\
Project & All hunks & \multicolumn{2}{r}{with single-statement pairs} \\
\midrule
ambari & 13,701 & 8,565 & (62.5\%) \\
camel & 43,237 & 28,672 & (66.3\%) \\
hadoop & 30,806 & 21,049 & (68.3\%) \\
jetty & 38,443 & 25,517 & (66.4\%) \\
wicket & 32,132 & 21,926 & (68.2\%) \\
\bottomrule
\end{tabular}
\end{table}

\begin{table*}[!t]
\renewcommand{\arraystretch}{1.3}
\caption{Filtering results for training data.}
\label{tab:train}
\centering
\begin{tabular}{lrrrrrrrrrr}
\toprule
& \multicolumn{2}{c}{ambari} & \multicolumn{2}{c}{camel} & \multicolumn{2}{c}{hadoop} & \multicolumn{2}{c}{jetty} & \multicolumn{2}{c}{wicket} \\
\midrule
train period & \multicolumn{2}{c}{2011-2013} & \multicolumn{2}{c}{2007-2013} & \multicolumn{2}{c}{2009-2012} & \multicolumn{2}{c}{2009-2013} & \multicolumn{2}{c}{2004-2014} \\
\midrule
\#, (\%) of statement pairs \\
\cmidrule{1-1}
before filtering & 4,253 & (100\%) & 25,562 & (100\%) & 11,952 & (100\%) & 21,278 & (100\%) & 21,268 & (100\%) \\
(3) $<$3 tokens & 70 & (1.7\%) & 285 & (1.1\%) & 69 & (0.6\%) & 281 & (1.3\%) & 127 & (0.6\%) \\
(5) not parsable & 261 & (6.2\%) & 2,924 & (11.4\%) & 1,239 & (10.4\%) & 2,023 & (9.5\%) & 3,092 & (14.5\%) \\
(6) lost candidates & 88 & (2.1\%) & 1,489 & (5.8\%) & 384 & (3.2\%) & 1,186 & (5.6\%) & 1,324 & (6.2\%) \\
(7) identical & 2,353 & (55.3\%) & 11,346 & (44.4\%) & 5,812 & (48.6\%) & 7,755 & (36.4\%) & 7,068 & (33.2\$) \\
\rowcolor{gray!20} final statement pairs & 1,481 & (34.8\%) & 9,518 & (37.2\%) & 4,448 & (37.2\%) & 10,033 & (47.2\%) & 9,657 & (45.4\%) \\
\bottomrule
\end{tabular}
\end{table*}

Figure~\ref{fig:statement_distribution} shows the distribution of the number of pre- and post-correction statements in all change hunks (counted separately). 
We find that most of changes are single statements in either insertion, deletion, or modification. Multi-statement changes are not frequent. 
Table~\ref{tab:stat} shows the number of all change hunks and the number of change hunks that are derived from single-statement changes. We see from the table that approximately 62 -- 68\% of the changes are single-statement changes.
Since we investigated changes per methods using historage repositories~\cite{Hata:2011:HFV:2024445.2024463}, we could divide large modifications in files~\cite{4556130,4755633} to fine-grained changes, which results in high rations of single-statement changes.
These ratios are encouraging for \system, which is limited to single-statement changes.

\subsection{Experimental Design}

From the collected pre- and post-correction statements, we prepare the training data (Table~\ref{tab:train}) and testing data (Table~\ref{tab:test}). Considering the number of statements, we set the testing year for each project as shown in Table~\ref{tab:test}. All statement pairs in each testing year are used as testing data, which means we chose statement pairs whose pre-correction statements are created in the testing year and changed to the corresponding post-correction statements in the same testing year. All years before the testing year are considered as training periods. In each training period, the numbers of statement pairs, whose pre-correction statements are changed to post-correction statements in the training period, are shown in Table~\ref{tab:train}.

This experimental design can be regarded as a simulation of generating corrected statements only by learning past histories when new statements are created and they will be modified soon (in the same year). If this works, we can prevent recurring or similar issues before being inserted into the code, or even when the code is being edited.
For this purpose, we prepare the training and testing data by considering chronological order instead of random partitioning. For the risk of increasing unseen changes in the training data, we limit the testing year to one year.

\subsection{Data Preparation}
\label{ssec:data}

Table~\ref{tab:train} details the impact of the various preprocessing steps on our approach. The before filtering row shows the number of all single-statement change pairs. The $<3$ tokens row shows the effect of removing statements that have less than 3 tokens. Then we remove the unparsable statements in both, pre-correction and post-correction statements. The final step removes identical statement pairs in the pre- and post-correction statements. The last row shows the final number of statements used in our study.

In addition, we perform specific processing for the training and testing data, which we detail below:

\noindent \textbf{Replacing rare tokens in the training data.}
From the processed statement pairs, we prepare pre-correction statement corpus and post-correction statement corpus. For each corpus, tokens that appear only once are replaced with $\langle$unk$\rangle$, which is a common way to handle unknown tokens~\cite{neubig2017neural}.
This script is available in the \textsf{lamtram} toolkit.\footnote{\textsf{lamtram}: \url{https://github.com/neubig/lamtram}}

\noindent \textbf{Categorization of testing data.}
When testing our approach, we call the pre-correction statements in the testing data as \textit{queries}. On the other hand, we call the post-correction statements as \textit{references}. 

When we evaluate our approach, we separate the testing data with their characteristics. First, all statement pairs in the testing data are divided into bug-fix statement pairs and non-bug-fix statement pairs. This classification procedure is presented in Section~\ref{ssec:guru}. Then both classes of statement pairs are categorized into three:

\begin{description}
\item[NU:] \textbf{No unknown.} There are no unknown tokens in a statement pair. All tokens in a query statement appear in the pre-correction statement from the training data corpus, and all tokens in a reference statement appear in the post-correction statement of the training data corpus.

\item[UQ:] \textbf{Unknown in query.} One or more token(s) in the query statement do not appear in the pre-correction statement corpus. In other words, there are unknown tokens in the query.

\item[UR:] \textbf{Unknown in reference.} Although there is no unknown token in the query statement, there are one or more unknown token(s) in the reference, i.e., in the corresponding post-correction statement.
\end{description}

We categorize the statements as shown above to know which data can be used in our experiments. This is particularly important since the trained NMT models have not seen \textit{unknown} tokens during training, addressing queries in UQ or UR is very difficult. In fact, it is impossible for our model to generate statements that are the exact same as the references for the UR category.%
\footnote{These errors could potentially be alleviated by incorporating a mechanism to copy inputs from the source \cite{gu-EtAl:2016:P16-1} or generate tokens in several sub-token parts \cite{sennrich-etal-2016-neural}.}

\begin{table}[!t]
\renewcommand{\arraystretch}{1.3}
\caption{Summary of testing data.}
\label{tab:test}
\centering
\begin{tabular}{lrrrrr}
\toprule
& ambari & camel & hadoop & jettty & wicket \\
\midrule
test year & 2014 & 2014 & 2013 & 2014 & 2015 \\
\cmidrule{1-1}
\rowcolor{gray!20} NU & 6 & 42 & 29 & 149 & 7 \\
UQ & 30 & 54 & 167 & 31 & 16 \\
UR & 0 & 8 & 15 & 13 & 1 \\
\bottomrule
\end{tabular}
\end{table}

Table~\ref{tab:test} shows the number of statement pairs for these categories of bug-fixing and non-bug fixing classes. As can be seen from the Table~\ref{tab:test}, the majority of the training data's statements fall in the UQ category (except for the Jetty project). On the other hand, the good news is that statements in the UR category are the least. We evaluate our approach using statements in the NU category. 


The dataset is available online.\footnote{Train and test dataset for this study: \dataset} 

\subsection{Identifying Bug-Fixing Statements}
\label{ssec:guru}
We collect bug-fixing statement pairs by identifying the pairs of bug-inducing and bug-fixing commits. To obtain these commits, we use \texttt{Commit.guru}~\cite{Rosen:2015:CGA:2786805.2803183}, a tool that analyzes and provides change level analytics.\footnote{Commit Guru: \url{http://commit.guru}} For full details about commit.guru, we point the reader to the paper by Rosen \emph{et al.}~\cite{Rosen:2015:CGA:2786805.2803183}, however, here we describe the relevant details for our paper.
Commit.guru takes as input a Git repository address, an original code repository in this study, and provides data for all commits of the project. It applies the SZZ algorithm~\cite{Sliwerski:2005:CIF:1083142.1083147} to identify bug-inducing commits and their associated bug-fixing commits. In addition, Commit.guru provides a number of change level metrics related to the size of the change, the history of the files changed, the diffusion of the change and the experience of the developers making the modification. 

As mentioned earlier in step (1) of preprocessing (Section~\ref{ssec:mining}), we have meta information of statements including the original commits of post-correction statements and the original commits of pre-correction statements.
We consider a pair of statements bug fixing if and only if a pre-correction statement is created in a bug-inducing commit and an post-correction statement is created in the associated bug-fixing commit.\footnote{Commit IDs in a \textsf{historage} and the corresponding commit IDs in the original Git repository are different because the contents are different. But we can trace the corresponding original commit IDs from \textsf{historage} since they are written in \texttt{git notes} of \textsf{historage}.}
The other statement pairs are treated as non-bug-fix statements.
We do not distinguish the types of the training data, that is, bug-fix or non-bug-fix. This is because we prefer to increase the training data available to the model and make the model learn from all varieties of changes.

\section{Evaluation}
\label{sec:eval}

We evaluate the performance of \system~with respect to two aspects: accuracy and usefulness of generated statements. In all of the results presented in this section, the NMT models are trained and tested with data from the same project (i.e., within-project evaluation).

\subsection*{Can the models generate valid statements?}

\begin{table}[!t]
\renewcommand{\arraystretch}{1.3}
\caption{Number of the generated valid statements for \textit{buggy} queries.}
\label{tab:parse}
\centering
\begin{tabular}{ccccc}
\toprule
ambari & camel & hadoop & jetty & wicket \\
\midrule
6/6 & 42/42 & 28/29 & 147/149 & 7/7 \\
(100\%) & (100\%) & (97\%) & (99\%) & (100\%) \\
\bottomrule
\end{tabular}
\end{table}

%
%

\begin{figure*}[!t]
\centering
\subfloat[ambari]{
\begin{tabular}{c}
\includegraphics[width=0.16\linewidth]{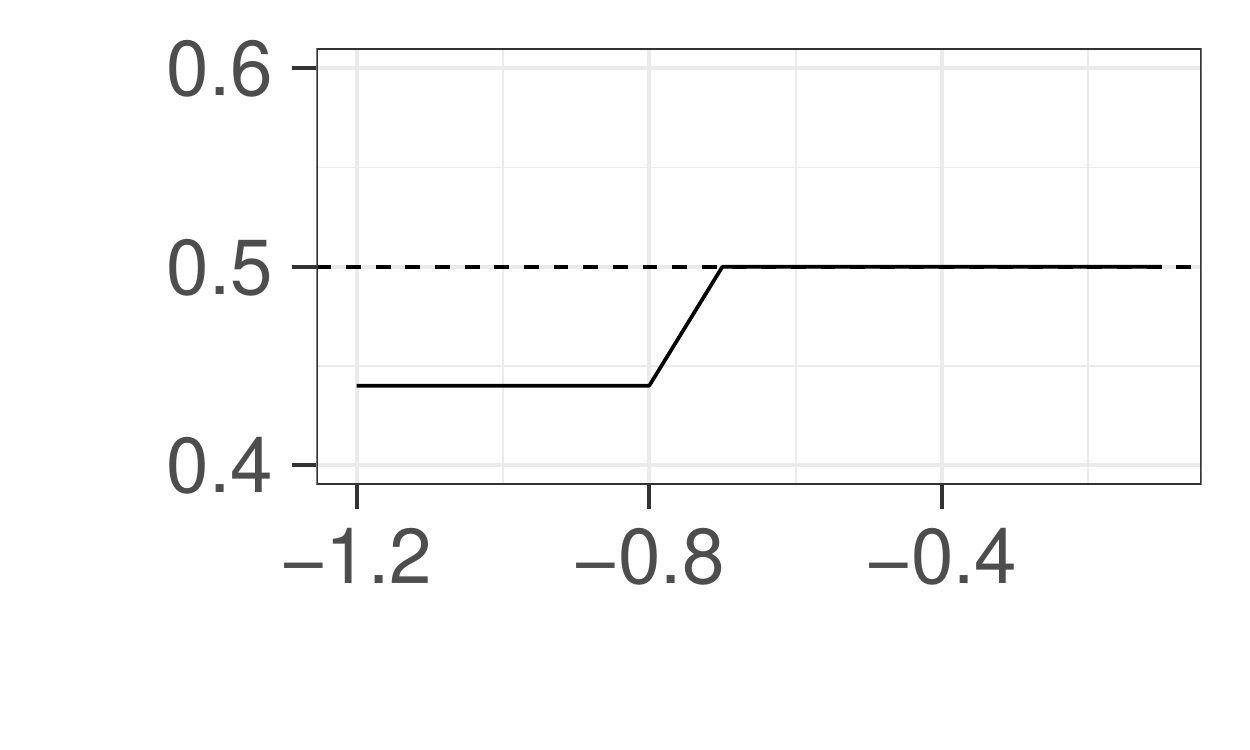} 
\end{tabular}
}%
\label{fig:sfig1}
\hfil
\subfloat[camel]{
\begin{tabular}{c}
\includegraphics[width=0.16\linewidth]{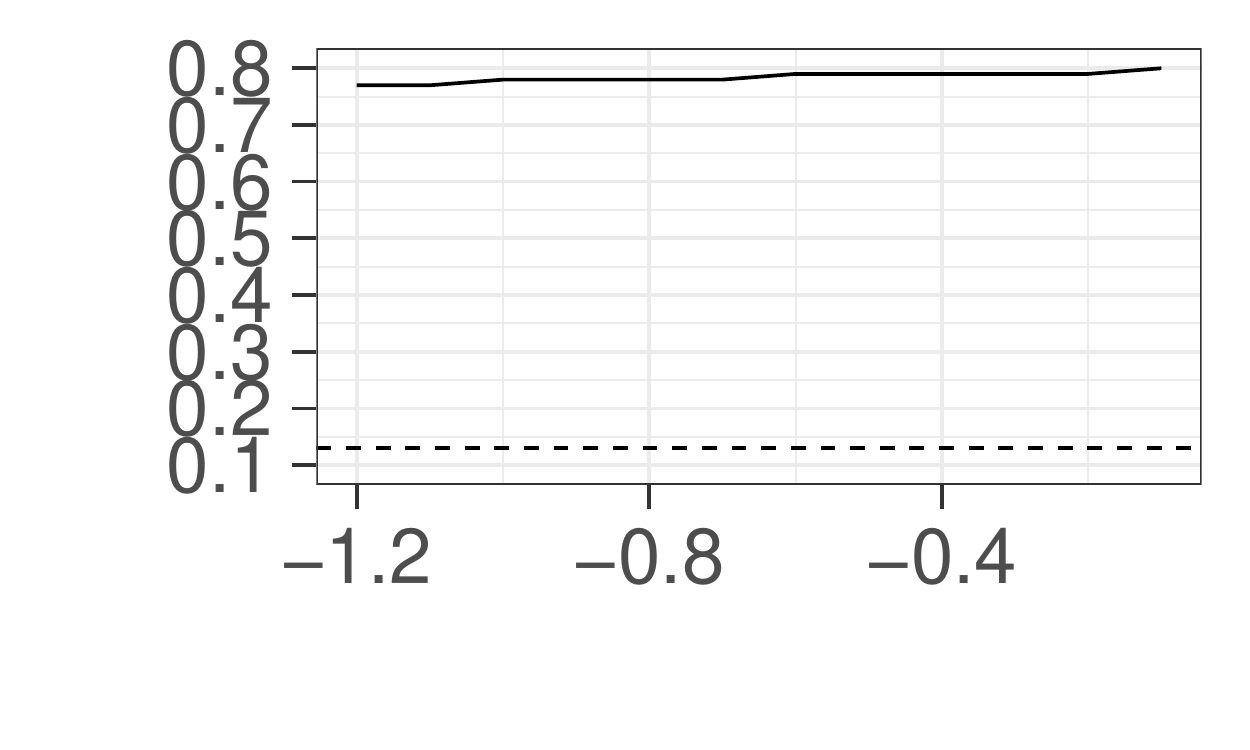} 
\end{tabular}
}%
\label{fig:sfig2}
\hfil
\subfloat[hadoop]{
\begin{tabular}{c}
\includegraphics[width=0.16\linewidth]{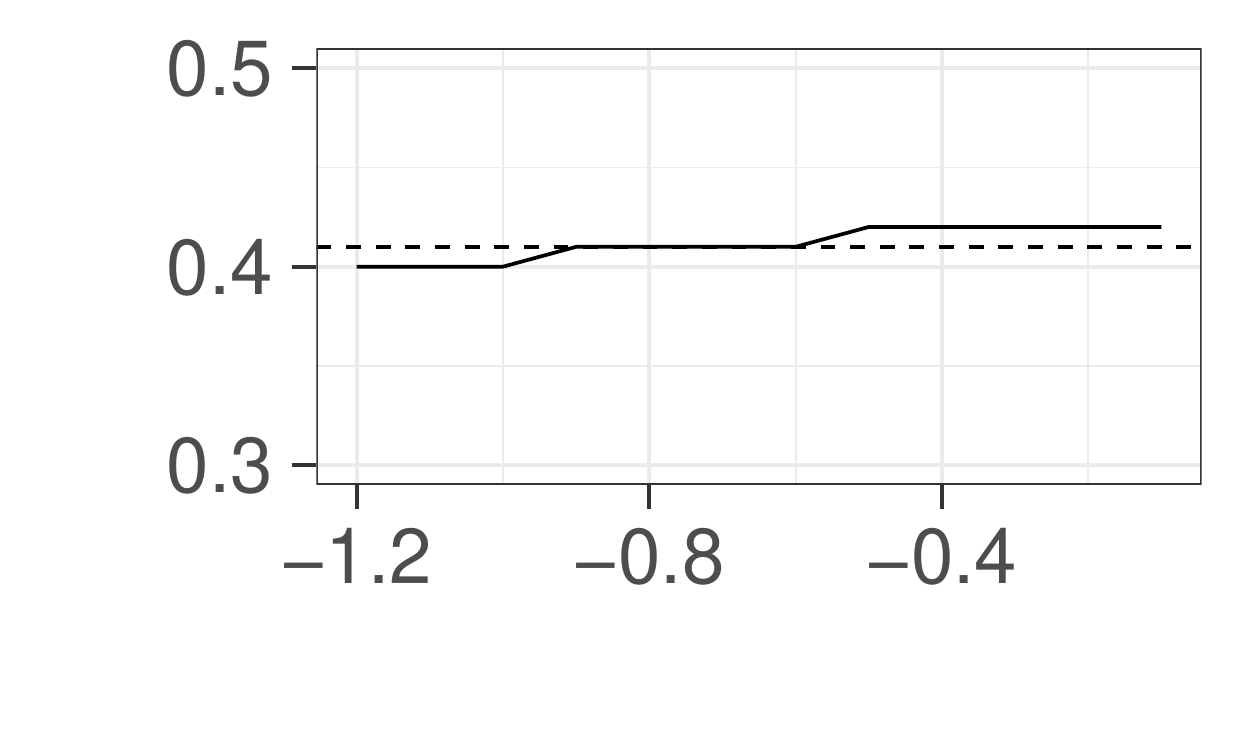} 
\end{tabular}
}%
\label{fig:sfig3}
\hfil
\subfloat[jetty]{
\begin{tabular}{c}
\includegraphics[width=0.16\linewidth]{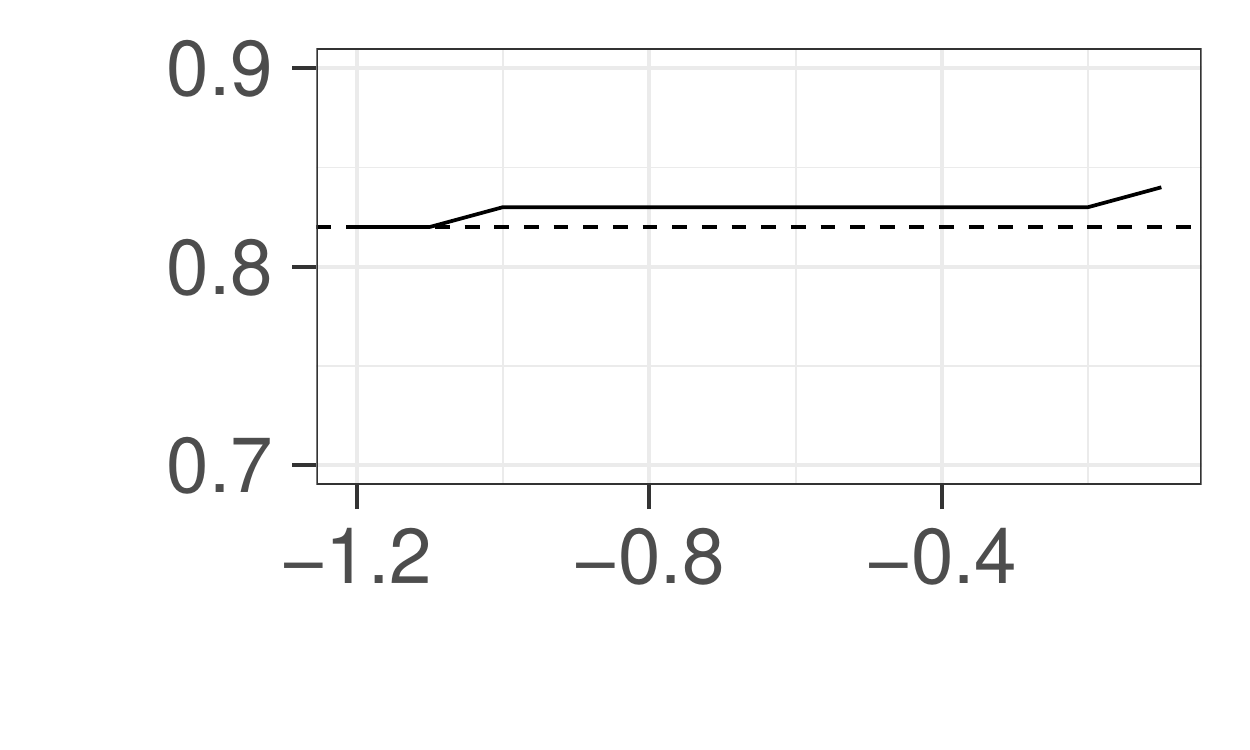} 
\end{tabular}
}%
\label{fig:sfig4}
\hfil
\subfloat[wicket]{
\begin{tabular}{c}
\includegraphics[width=0.16\linewidth]{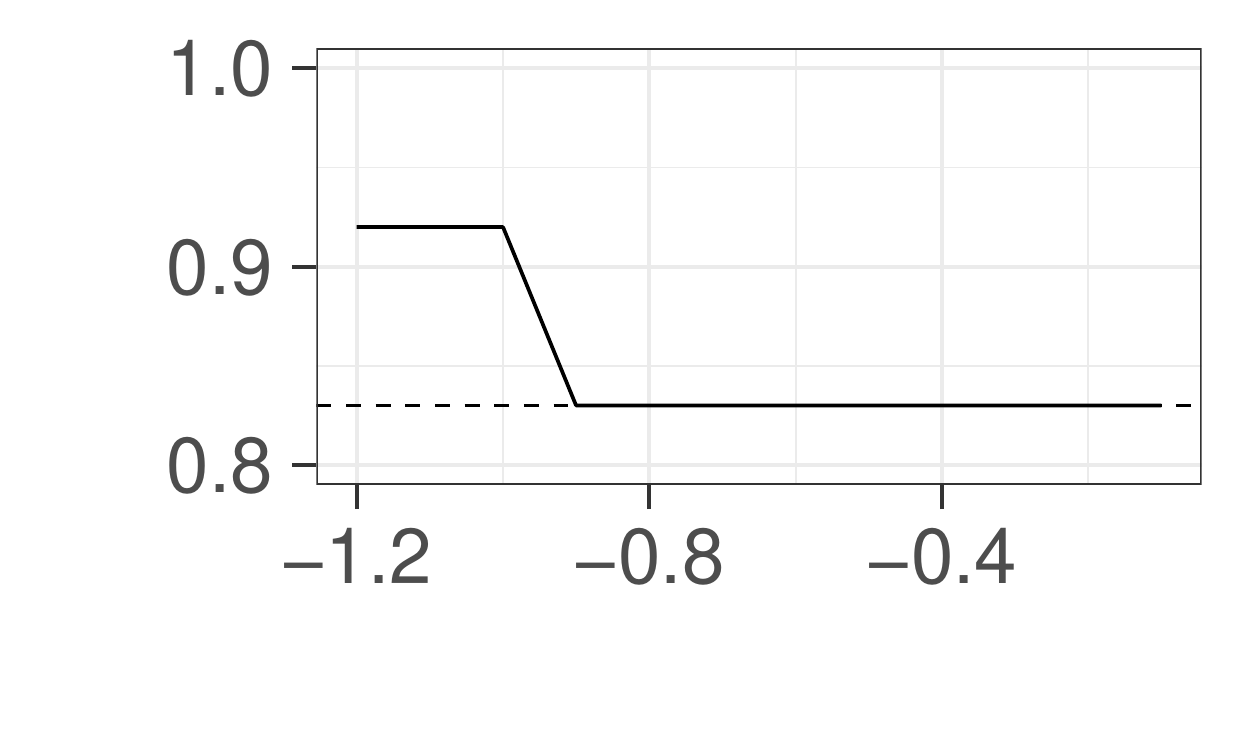} 
\end{tabular}
}%
\label{fig:sfig5}
\caption{$F_1$ values with thresholds (-1.2 to -0.1). The solid lines are $F_1$ values of \system~and the dotted lines are $F_1$ values of the baseline.}
\label{fig:nbest}
\end{figure*}

We consider complete and parsable statements as valid statements. We investigate whether the generated statements are valid using the same process of step (5) in the preprocessing described in Section~\ref{ssec:pre}.
Table~\ref{tab:parse} shows the number (and percentage) of generated valid statements. We do not use thresholds here, that is, all generated statements including low scores are considered. As we see from the table, in most cases the models generated valid and complete Java statements. 
These high accuracy results are especially interesting since we did not explicitly teach the models the Java language specification. Simply, the models were able to achieve this high level of performance by themselves, using approx. 1,500 to 10,000 statement pairs.

\begin{tcolorbox}
In most cases of the five projects nearly 100\% of the generated statements are valid Java statements. In total the models generated 230 valid statements for 233 queries (98.7\%).
\end{tcolorbox}

\subsection*{How accurate are the generated statements?}


In this section we evaluate the accuracy in a strict manner, that is, only generated statements that are \textbf{identical to references} are considered as correct. Our results are based on the NU (no unknown) category of statements, since as mentioned earlier, other categories are difficult (impossible for UR) to generate accurate statements that are identical to the reference statements.

Before analyzing accuracy, we categorize the outputs into four types:
\begin{itemize}
\item \textbf{Correct:} a generated statement is identical to the reference, including arguments.
\item \textbf{Argument incorrect:} a generated statement is identical to the referecne, except for the arguments.
\item \textbf{Incorrect:} a generated statement is not identical to the reference, even if we exclude the arguments.
\item \textbf{NA:} a generated statement is invalid or identical to the query, or its score is lower than a threshold.
\end{itemize}


To measure the accuracy of generated results, we compute precision, recall, and $F_1$, which are defined as:
$precision = \frac{\#correct}{\#provided}$, $recall = \frac{\%correct}{\#queries}$, and $F_1 = \frac{2 \times precision \times recall}{precision + recall}$, where $\#provided$ is the sum of $\#correct$, $\#argument$ $incorrect$, and $\#incorrect$. Higher precision indicates that the provided results are correct. Higher recall means that the results contain less NA but many correct. Providing a small number of results with high confidence can improve precision but lower recall. Since there is a tradeoff between precision and recall, $F_1$, the harmonic mean of precision and recall, is also presented.

We compare the accuracy of our NMT models with the pattern-based patch suggestion approach, i.e., patch suggestion with line-granular snippets that already exist in the training data, which serves as our baseline~\cite{Barr:2014:PSH:2635868.2635898}.
To do so, we examine whether a query statement exists in the pre-correction statement training data corpus, and if it does, we check whether the corresponding post-correction statement also exists (in the training data). If this happens, then we consider the statement to be covered by the Plastic Surgery approach. If there is no identical pre-correction statement, we mark the result as NA.
Note that this baseline is slightly different from the Plastic Surgery hypothesis since it searches post-correction statements from all packages, which requires the existence of pre-correction statements in the codebase. We apply the same argument replacing processing in Section~\ref{ssec:post} to make it fair.

\def\mybar#1{
  {\rule{#1mm}{5pt}}}

\begin{table}[!t]
\renewcommand{\arraystretch}{1.3}
\caption{Fix generation for \textit{buggy} queries. Threshold is -0.7. Bold indicates win in comparison with the baseline.}
\label{tab:buggy}
\centering
\begin{tabular}{llrlrrlr}
\toprule
&  & \multicolumn{3}{c}{\system} & \multicolumn{3}{c}{Baseline} \\
\midrule
amb. & correct & 2 & \mybar{3.3} & (33\%) & 2 & \mybar{3.3} & (33\%) \\
& arg incor. & 0 & \mybar{0} & (0\%) & 0 & \mybar{0} & (0\%) \\
& incor. & 0 & \mybar{0} & (0\%) & 0 & \mybar{0} & (0\%) \\
& NA & 4 & \mybar{6.7} & (67\%) & 4 & \mybar{6.7} & (67\%) \\
\cmidrule{2-8}
& Pr, Re, F & 1.00 & 0.33 & 0.50 & 1.00 & 0.33 & 0.50 \\
\midrule
cam. & correct & 26 & \mybar{6.2} & (62\%) & 1 & \mybar{0.2} & (2\%) \\
& arg incor. & 2 & \mybar{0.5} & (5\%) & 2 & \mybar{0.5} & (5\%) \\
& incor. & 2 & \mybar{0.5} & (5\%) & 1 & \mybar{0.2} & (2\%) \\
& NA & 12 & \mybar{2.9} & (29\%) & 38 & \mybar{9} & (90\%) \\
\cmidrule{2-8}
& Pr, Re, F & \textbf{0.87} & \textbf{0.62} & \textbf{0.72} & 0.25 & 0.02 & 0.04 \\
\midrule
had. & correct & 7 & \mybar{2.4} & (24\%) & 7 & \mybar{2.4} & (24\%) \\
& arg incor. & 3 & \mybar{1.0} & (10\%) & 3 & \mybar{1.0} & (10\%) \\
& incor. & 10 & \mybar{3.4} & (34\%) & 10 & \mybar{3.4} & (34\%) \\
& NA & 9 & \mybar{3.1} & (31\%) & 9 & \mybar{3.1} & (31\%) \\
\cmidrule{2-8}
& Pr, Re, F & 0.35 & 0.24 & 0.29 & 0.35 & 0.24 & 0.29 \\
\midrule
jet. & correct & 101 & \mybar{6.8} & (68\%) & 101 & \mybar{6.8} & (68\%) \\
& arg incor. & 11 & \mybar{0.7} & (7\%) & 11 & \mybar{0.7} & (7\%) \\
& incor. & 9 & \mybar{0.6} & (6\%) & 11 & \mybar{0.7} & (7\%) \\
& NA & 28 & \mybar{1.9} & (19\%) & 26 & \mybar{1.7} & (17\%) \\
\cmidrule{2-8}
& Pr, Re, F & \textbf{0.83} & 0.68 & \textbf{0.75} & 0.82 & 0.68 & 0.74 \\
\midrule
wic. & correct & 5 & \mybar{8.1} & (71\%) & 5 & \mybar{8.1} & (71\%) \\
& arg incor. & 0 & \mybar{0} & (0\%) & 0 & \mybar{0} & (0\%) \\
& incor. & 0 & \mybar{0} & (0\%) & 0 & \mybar{0} & (0\%) \\
& NA & 2 & \mybar{2.9} & (29\%) & 2 & \mybar{2.9} & (29\%) \\
\cmidrule{2-8}
& Pr, Re, F & 1.00 & 0.71 & 0.83 & 1.00 & 0.71 & 0.83 \\
\bottomrule
\end{tabular}
\end{table}

When evaluating \system~and as stated in Section~\ref{ssec:post}, we use a threshold to ignore results with low confidence. Figure~\ref{fig:nbest} illustrates the transitions of $F_1$ values with different thresholds (from -1.2 to -0.1).
The solid lines are $F_1$ values of \system~and the dotted lines are $F_1$ values of the baseline, which do not change with thresholds. We find that $F_1$ values slightly change when we vary the thresholds. Lowering thresholds improves recall, however, it impacts the precision in the opposite direction. On the other hand, raising the threshold improves precision but makes recall worse.\footnote{The data of wicket is an exception, in which we can find a correct result without increasing incorrect statements when lowering the threshold.}
Based on our analysis of the threshold, we empirically set the threshold as -0.7 for the analyses that follow.

\begin{table*}[!t]
\renewcommand{\arraystretch}{1.3}
\caption{Examples of generated statements that cannot be fixed by the baseline.}
\label{tab:example}
\centering
\begin{tabularx}{\linewidth} {lXl}
\toprule
& Query statement & Generated statement \\
\midrule
1. & commands [ 10 ] = this . passwordFile . toString ( ) ; & commands [ 11 ] = this . passwordFile . toString ( ) ; \\
2. & List body = assertIsInstanceOf ( arg$\dagger$ ) ; & List $< ? >$ body = assertIsInstanceOf ( arg$\dagger$ ) ; \\
3. & Set $<$ String $>$ knownRoles = new HashSet ( ) ; & Set $<$ String $>$ knownRoles = new HashSet $<>$ ( ) ; \\
4. & return this . height ; & return height ; \\
\bottomrule
\multicolumn{2}{l}{arg$\dagger$: List . class , result . getExchanges ( ) . get ( 0 ) . getIn ( ) . getBody ( )} \\
\end{tabularx}
\end{table*}

Table~\ref{tab:buggy} 
shows the results of our approach and compares it with the results of the baseline. We observe that, as reported in the Plastic Surgery hypothesis paper~\cite{Barr:2014:PSH:2635868.2635898}, the baseline of the pattern-based patch recommendation works in many cases, that is, changes (corrections) contain snippets that already exist in code repositories at the time of the changes, and these snippets can be efficiently found and exploited.
That said, Table~\ref{tab:buggy} shows that \system~improves the results in two projects and does not change in three projects. We observe that in camel, the results are greatly improved: 26 correct statements are generated compared with one correct recommendation from the pattern matching of the baseline. 
Our results show that the NMT models work, as well as the baseline, if there are easily exploited statement-level patterns (i.e., reusable snippets), and works better than the baseline if there exist only finer-grained exploited patterns (i.e., fine-grained fixing patterns), which the statement-based pattern matching cannot use.


Table~\ref{tab:example} presents examples of generated fixes that cannot be fixed by the baseline, but have a fix generated with our models. Sometimes the model learns the incrementation of value (query 1). Generics-related fixes are typical examples of successful generation with the NMT models (query 2 and 3). Sometimes it is preferred to remove \texttt{this} (query 4) if it makes the style consistent with the styling used in the specific project. In fact, our models learned to remove the keyword `this' because similar patterns were prevalent in the project's history.

\begin{tcolorbox}
NMT-based patch generation works better than pattern-based patch suggestion, achieving $F_1$ values between 0.29 to 0.83 for buggy queries. 
In total 157 correctly generated statements without method arguments, the contents of method arguments for 141 statements (89.8\%) are correctly provided by reusing the contents of method arguments in queries.
\end{tcolorbox}

\subsection*{Do humans detect similarity between generated statements and actual statements?}

During the previous evaluation of accuracy, we considered the generated statements to be correct only if they are \emph{identical to the reference statements}, otherwise they are considered to be incorrect or NA. To investigate whether the generated statements are useful, even if they are not identical to actual future corrections, we also perform a human evaluation with such (non-identical) corrections.

We show survey participants the following three code snippets for one fix: i) an original problematic code snippet (before correction), ii) the actually fixed code snippet (after correction), and iii) a code snippet that is proposed as a fix by our NMT models. 
All code snippets contain one type of buggy or fixed statements with the surrounding statements.\footnote{One case has two buggy or fixed statements that are similar to each other, and others only have one statement of buggy or fixed statement.}

From the five projects, we collect ten corrections including five correctly and five incorrectly generated statements in the NU (no unknown) category, which are evaluated in Table~\ref{tab:buggy}. 
In addition, we collect five fixes that belong to the UQ (unknown in query) or the UR (unknown in reference) categories, which are known to be difficult for NMT models to generate. For simplicity, we call the above three groups \textit{correct fixes}, \textit{incorrect fixes}, and \textit{challenging fixes} respectively.

For each correction we prepare the following four statements, and ask the participants to evaluate using a five-level Likert scale scores from 1 (strongly disagree) to 5 (strongly agree) whether:
(a) \textit{The proposed fix helps you to understand the required change},
(b) \textit{The proposed fix can be a reference if you were to create your own fix},
(c) \textit{The proposed fix is harmful or confusing}, and
(d) \textit{The proposed fix does not make sense and I will just ignore it}.
We asked not only positive impressions but also negative impressions in order to assess the usefulness and potential risks of incorrect generation.
The survey material is available online.\footnote{Survey material for human evaluation: \url{https://tinyurl.com/RachetSurvey}} 

We recruited participants in Canada, US, and Japan, and 20 people participated in the survey including five undergraduate, 14 graduate students, and one professor.
As Siegmund \emph{et al.} reported that self estimation seems to be a reliable way to measure programming experience~\cite{Siegmund:2014:MMP:2674501.2674544}, we asked the participants to estimate their experience in both, overall and Java programming experience. The participants can select any of 5 choices, varying between 1 (very inexperienced) to 5 (very experienced). Those who score 4 or 5 in both self estimation are considered to be \textit{high}-experienced and others are considered to have \textit{low}-experience.
Five in six high experienced participants have more than five years of development experience, and the other have three-to-five years of experience. In 14 low-experience participants, the experience periods vary from less than one year, one-to-three years, three-to-five years, and more than five years.


\begin{figure}[!t]
\centering
\includegraphics[width=\linewidth]{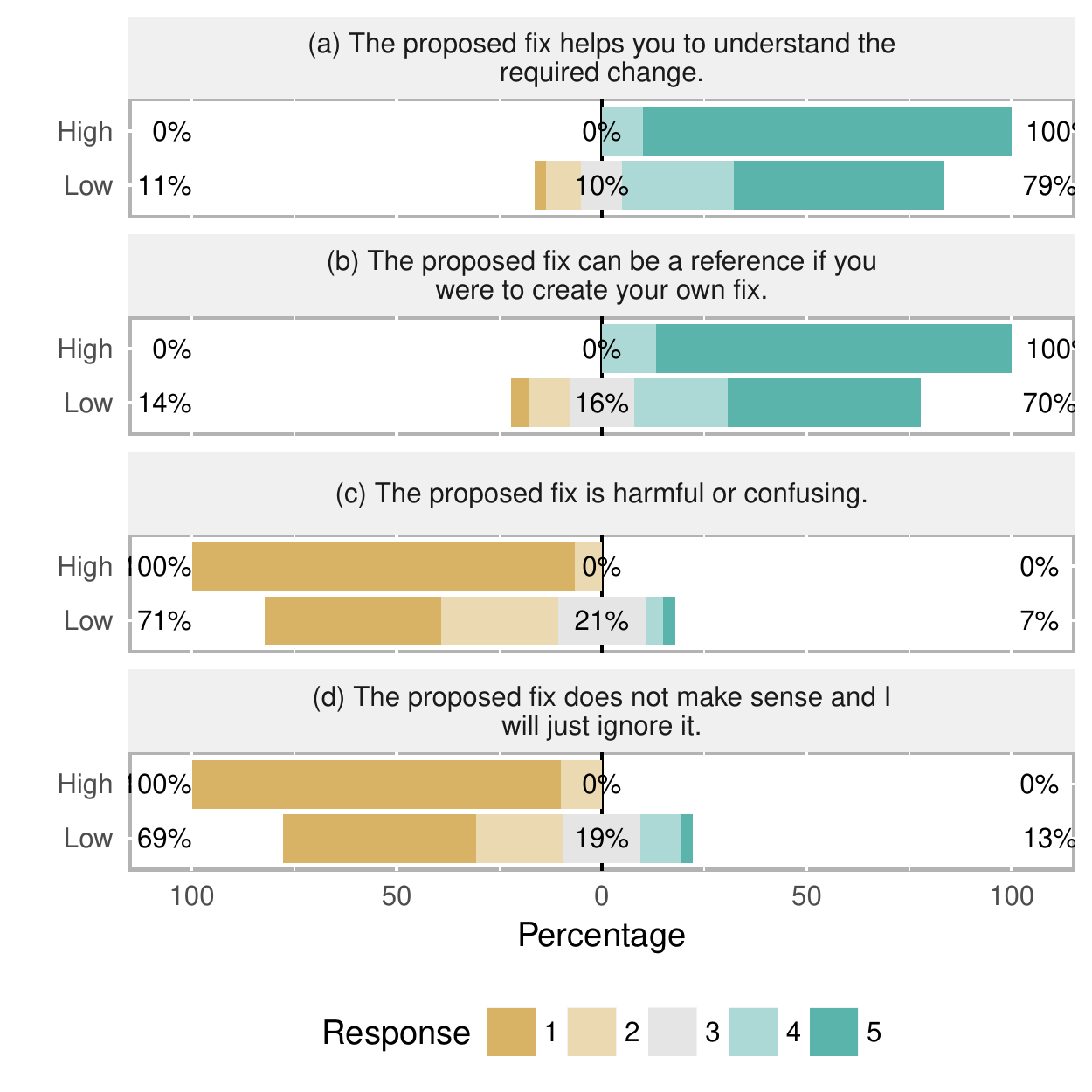}
\caption{Survey results of five \textit{correct fixes}. Responses are Likert scale from 1 (strongly disagree) to 5 (strongly agree).}
\label{fig:correct}
\end{figure}

\begin{figure}[!t]
\centering
\includegraphics[width=\linewidth]{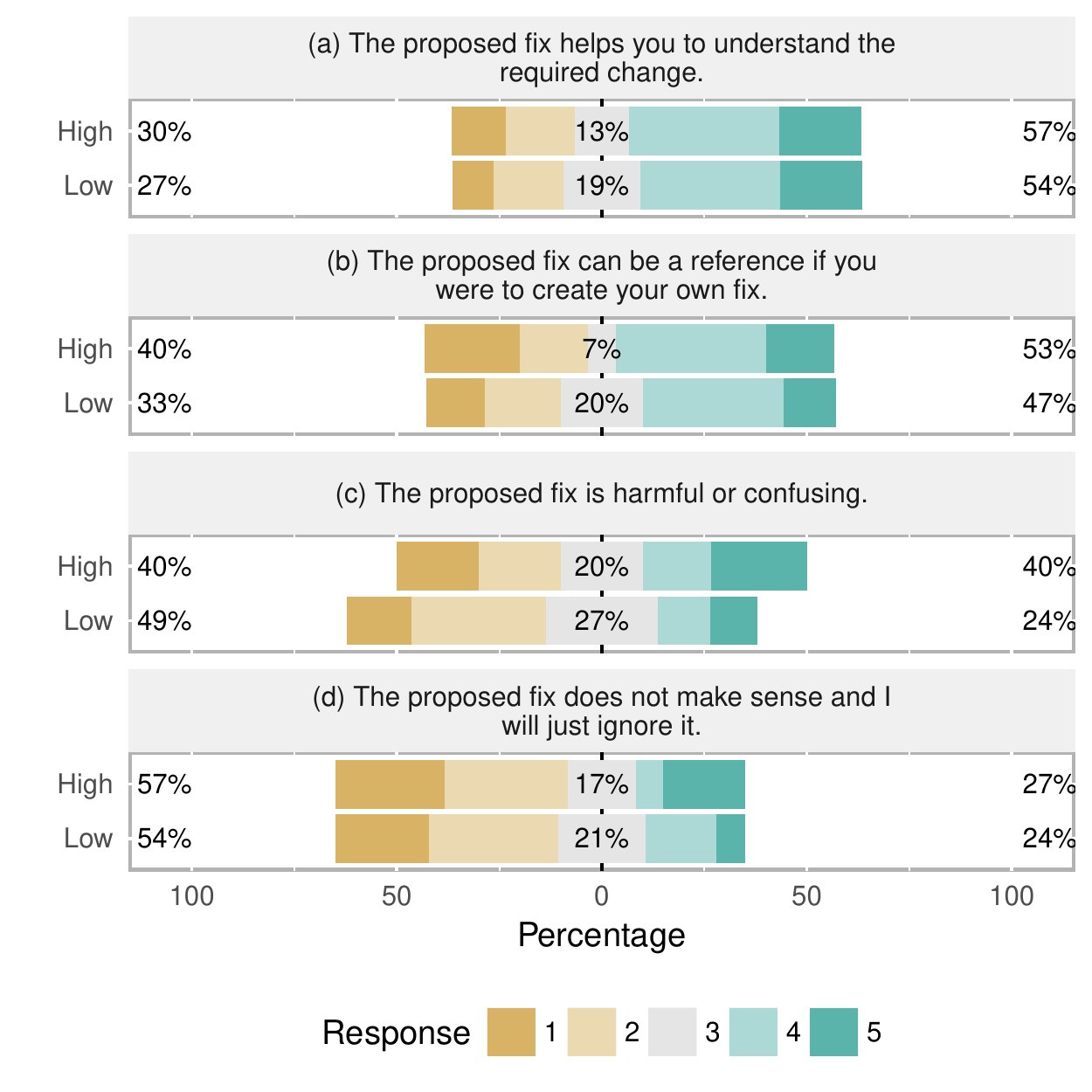}
\caption{Survey result of five \textit{incorrect fixes} with six high and 14 low experienced participants.}
\label{fig:incorrect}
\end{figure}

\begin{figure}[!t]
\centering
\includegraphics[width=\linewidth]{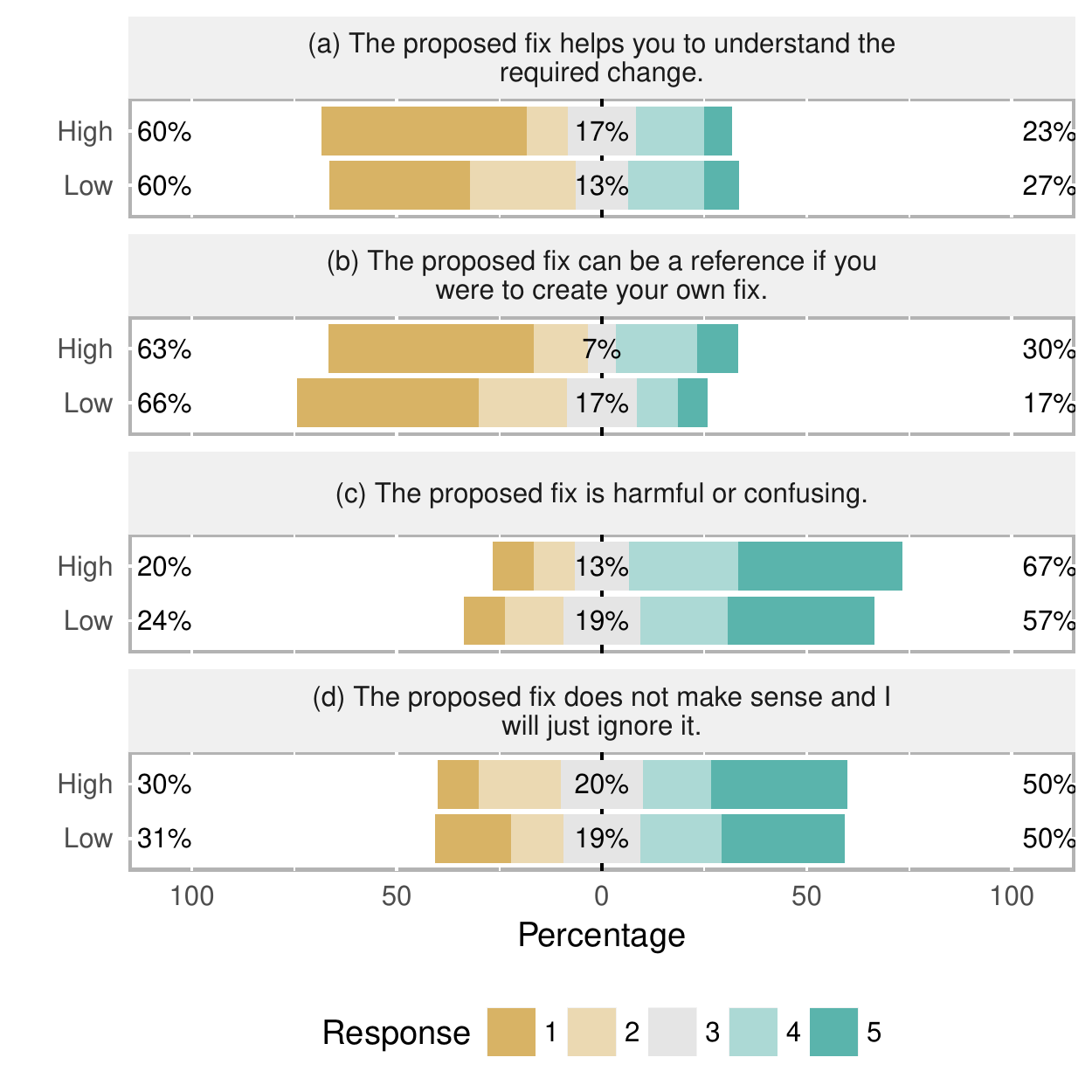}
\caption{Survey result of five \textit{challenging fixes} with six high and 14 low experienced participants.}
\label{fig:challenge}
\end{figure}


Figure~\ref{fig:correct} shows the result of the \textit{correct fix} group. The results shown in the figure show that the generated statements are useful. All high-experience and most of the low-experience participants agreed (scores 4 or 5 for questions (a) and (b)) that the correct fix statements helped them and that the statements and did not have negative effects (i.e., most scores are 1 or 2 for questions (c) and (d)). 


Figure~\ref{fig:incorrect} shows the results of the \textit{incorrect fix} group, which includes statements with incorrect method calls and/or incorrect generic types, for example. We assumed that these fixes are harmful or confusing because they tend to be partially the same as the references, but are slightly different from the actual fixes. However, as evidenced by the results shown in Figure~\ref{fig:incorrect}, the majority of highly and low experienced participants agree to that such imperfect statements may still be helpful (i.e., by providing positive answers to statement (a) and (b)). Although the highly experienced participants tend to consider such imperfect statements harmful or confusing (highly experienced agreed 40\% and low experienced agreed 24\%), both high and low experienced participants did not consider the proposed fix did not make sense. 


The following are some comments we received: ``\textit{A potentially better fix than original},'' ``\textit{I prefer having a `this' but this is personal preference.},'' ``\textit{The word `info' seems more clear than `trace.' Good change},'' and ``\textit{Changed to a wrong direction}. One participant pointed out that the  proposed fixes seem to provide several pieces of information, for example, the location of fix, the need of initialization of methods, and types for generics. S/he claimed that this information is useful if s/he knows the context of the code, even if an error exists. We find that even for the same fixes, some participants perceived them differently, from which we can infer that sometimes better fixes depend on preferences and/or the context. 

Figure~\ref{fig:challenge} shows the result of the \textit{challenging fix} group. The fixes belonging to this group are difficult to generate because of unknown terms, which means that the queries and correct answers are mostly unseen by the models. Therefore we considered that those fixes did not make sense and did not provide any useful information. One case is changing \texttt{BigInteger} to \texttt{toHexString}, which even fails the compilation check. A participant left a comment ``\textit{I think it would produce more confusion than help.}''
As seen from the figure, the majority of both, highly and low experienced participants have negative impressions with regards to such statements. However, some still have positive opinions even for such failure cases.
Another case is given a query of `\texttt{FSDataOutputStream fos = null ;},' generating `\texttt{HdfsDataOutputStream copyError = null ;}' while the correct answer is `\texttt{HdfsDataOutputStream fos = null ;}.' This happens because the term `\texttt{fos}' does not appear in the target (i.e., post-correction statement) corpus. In the training corpora, there is no statement co-occurring \texttt{FSDataOutputStream} with \texttt{fos} or \texttt{null}. The model learned the replacement of \texttt{FSDataOutputStream} and \texttt{HdfsDataOutputStream} from the different context of statements.
For this case, the participants evaluated more positively than negatively, although there were some comments which stated that the generated statements can be confusing. In sum, we find that even for the challenging fixes, they might be useful.

\begin{tcolorbox}
Even if generated fixes are not identical to actual fixes, they can be helpful because they can suggest the locations of required changes and possible replacements/insertions/deletions. Sometimes better fixes depend on personal preferences or the styles of projects. Although NMT models can learn fine-grained patterns of changes, the lack of information or novel queries are major challenges of fix generation. 
\end{tcolorbox}

\section{Discussions}
\label{sec:discussion}
\begin{table}[!t]
\renewcommand{\arraystretch}{1.3}
\caption{Summary of testing data from not bug-fixing statements.}
\label{tab:test-nonbug}
\centering
\begin{tabular}{lrrrrr}
\toprule
& ambari & camel & hadoop & jettty & wicket \\
\midrule
test year & 2014 & 2014 & 2013 & 2014 & 2015 \\
\cmidrule{1-1}
\rowcolor{gray!20} NU & 64 & 114 & 135 & 151 & 28 \\
UQ & 792 & 386 & 1,274 & 428 & 99 \\
UR & 38 & 46 & 113 & 148 & 7 \\
\bottomrule
\end{tabular}
\end{table}

\begin{table}[!t]
\renewcommand{\arraystretch}{1.3}
\caption{Number of the generated valid statements for \textit{non-buggy} queries.}
\label{tab:parse2}
\centering
\begin{tabular}{ccccc}
\toprule
ambari & camel & hadoop & jetty & wicket \\
\midrule
63/64 & 113/114 & 134/135 & 149/151 & 20/28 \\
(98\%) & (99\%) & (99\%) & (99\%) & (71\%) \\
\bottomrule
\end{tabular}
\end{table}

\begin{table}[!t]
\renewcommand{\arraystretch}{1.3}
\caption{Fix generation for \textit{non-buggy} queries. Threshold is -0.7. Bold indicates win in comparison with the baseline.}
\label{tab:nonbuggy}
\centering
\begin{tabular}{llrlrrlr}
\toprule
& & \multicolumn{3}{c}{\system} & \multicolumn{3}{c}{Baseline} \\
\midrule
amb. & correct & 6 & \mybar{0.9} & (9\%) & 5 & \mybar{0.8} & (8\%) \\
& arg incor. & 7 & \mybar{1.1} & (11\%) & 6 & \mybar{0.9} & (9\%) \\
& incor. & 10 & \mybar{1.6} & (16\%) & 17 & \mybar{2.7} & (27\%) \\
& NA & 41 & \mybar{6.4} & (64\%) & 36 & \mybar{5.6} & (56\%) \\
\cmidrule{2-8}
& Pr, Re, F & \textbf{0.26} & \textbf{0.09} & \textbf{0.14} & 0.18 & 0.08 & 0.11 \\
\midrule
cam. & correct & 33 & \mybar{2.9} & (29\%) & 30 & \mybar{2.6} & (26\%) \\
& arg incor. & 3 & \mybar{0.3} & (3\%) & 4 & \mybar{0.4} & (4\%) \\
& incor. & 34 & \mybar{3.0} & (30\%) & 25 & \mybar{2.2} & (22\%) \\
& NA & 44 & \mybar{3.9} & (39\%) & 55 & \mybar{4.8} & (48\%) \\
\cmidrule{2-8}
& Pr, Re, F & 0.47 & \textbf{0.29} & \textbf{0.36} & \textbf{0.51} & 0.26 & 0.35 \\
\midrule
had. & correct & 7 & \mybar{0.5} & (5\%) & 7 & \mybar{0.5} & (5\%) \\
& arg incor. & 21 & \mybar{1.6} & (16\%) & 22 & \mybar{1.6} & (16\%) \\
& incor. & 31 & \mybar{2.3} & (23\%) & 28 & \mybar{2.1} & (21\%) \\
& NA & 76 & \mybar{5.6} & (56\%) & 78 & \mybar{5.8} & (58\%) \\
\cmidrule{2-8}
& Pr, Re, F & 0.12 & 0.05 & 0.07 & 0.12 & 0.05 & 0.07 \\
\midrule
jet. & correct & 15 & \mybar{1.0} & (10\%) & 14 & \mybar{0.9} & (9\%) \\
& arg incor. & 4 & \mybar{0.3} & (3\%) & 3 & \mybar{0.2} & (2\%) \\
& incor. & 32 & \mybar{2.1} & (21\%) & 24 & \mybar{1.6} & (16\%) \\
& NA & 100 & \mybar{6.6} & (66\%) & 110 & \mybar{7.3} & (73\%) \\
\cmidrule{2-8}
& Pr, Re, F & 0.29 & \textbf{0.10} & 0.15 & \textbf{0.34} & 0.09 & 0.15 \\
\midrule
wic. & correct & 10 & \mybar{3.6} & (36\%) & 0 & \mybar{0} & (0\%) \\
& arg incor. & 0 & \mybar{0} & (0\%) & 0 & \mybar{0} & (0\%) \\
& incor. & 3 & \mybar{1.1} & (11\%) & 3 & \mybar{1.1} & (11\%) \\
& NA & 15 & \mybar{5.4} & (54\%) & 25 & \mybar{8.9} & (89\%) \\
\cmidrule{2-8}
& Pr, Re, F & \textbf{0.77} & \textbf{0.36} & \textbf{0.49} & -- & -- & -- \\
\bottomrule
\end{tabular}
\end{table}

\subsection{Generating Non-Bug-Fixing Statements}
\label{ssec:nonbug}

For the accuracy evaluation in Section~\ref{sec:eval}, we only considered bug-fixing statements. Here we investigate the applicability of \system~in a more general context, i.e., for non-bug-fixing statements as well. In the same test year, we collected non-bug-fixing statements as shown in Table~\ref{tab:test-nonbug}. Again, we use a similar setup as we did for bug-fixing statement evaluations and compare the generated statements with the baseline.

Table~\ref{tab:parse2} shows the number of generated valid statements.
Similar to the result for buggy queries in Table~\ref{tab:parse}, most generated statements are valid Java statements.
Table~\ref{tab:nonbuggy} shows the results for non-bug-fixing statements. 
The $F_1$ values for non-bug-fixing queries ranges between 0.07 to 0.49. These $F_1$ values are lower than the results obtained for the bug-fixing queries shown in Table~\ref{tab:buggy}. That said, we still observe that in all five projects, \system~outperforms the baseline.
One possible explanation for the lower performance is the fact that there are relatively more UQ and UR statement pairs for non-bug-fixing datasets (as seen by comparing Table~\ref{tab:test} and Table~\ref{tab:test-nonbug}), which indicates the unique nature of non-bug-fix changes.

\begin{table}[!t]
\renewcommand{\arraystretch}{1.3}
\caption{Characteristics of fixes for the targeted bugs and summary of testing data from Defects4J.}
\label{tab:defects4j}
\centering
\begin{tabular}{lccccc}
\toprule
& closure & lang & math & time & mockito \\
\midrule
\# bugs & 151 & 53 & 83 & 20 & 23 \\
\cmidrule{2-6}
one line & 34 (23\%) &  7 (13\%) & 30 (36\%) & 3 (15\%) & 5 (22\%) \\
only add & 40 (26\%) & 26 (49\%) & 18 (22\%) & 7 (35\%) & 13 (57\%) \\
\midrule
\rowcolor{gray!20} NU & 21 & 3 & 19 & 5 & 0 \\
UQ & 8 & 2 & 13 & 1 & 0 \\
UR & 54 & 4 & 17 & 6 & 5 \\
\bottomrule
\end{tabular}
\end{table}

\subsection{Applying to Bug Dataset}
We also investigate the applicability of \system~on a commonly used bug dataset, Defects4J~\cite{Just:2014:DDE:2610384.2628055}.\footnote{We used the version 2.0.0 from \url{https://github.com/rjust/defects4j}.} The same steps shown in Section~\ref{ssec:mining}, are applied (we only targeted Git repositories), and bug-fixing modifications within methods are identified.

Table~\ref{tab:defects4j} presents the characteristics of bugs in the dataset.
We see that only 13\% to 36\% of bugs include single statement (one-line) changes for fixing, which are applicable for \system. From the other bugs, 22\% to 57\% bugs require only addition of statements, which is not targeted by our current approach.
From the bug-fixing commits, using the same steps detailed in Section~\ref{ssec:pre}, we collected pre- and post-correction statements for testing data.
Training data was prepared from the other commits.
As seen in Table~\ref{tab:defects4j}, there are 48 statements in the NU category.
Similar to other settings, 
we use the buggy statements as queries.

\begin{table}[!t]
\renewcommand{\arraystretch}{1.3}
\caption{Correct fix generations for the bugs in Defects4J.}
\label{tab:defects4jresult}
\centering
\begin{tabular}{lll}
\toprule
bug\_id & required fix & result \\
\midrule
closure-30 & 1 one-line and 1 large change & 1 statement with \\
& & incorrect arguments \\
closure-46 & 1 one-line & complete fix \\
& & including arguments \\
closure-134 & 4 add, 1 del., 4 one-line, and & 1 statement with \\
& 3 large changes & incorrect arguments \\
\bottomrule
\end{tabular}
\end{table}

\system~generated one complete fix and two partial fixes, which are shown in Table~\ref{tab:defects4jresult}.
The bug of closure-46 requires one single statement fix and \system~could successfully generate the fix including an argument. The other two bugs require multiple and larger changes, and \system~generated \textit{argument incorrect} fixes of single statements. Since the proposed approach does not account for the context of changes, it cannot handle appropriate argument changes.
Compared to 26 successful fixes by \textsf{ELIXIR}, one of the recent program repair techniques~\cite{Saha:2017:EEO:3155562.3155643}, the performance of \system~is low on the Defect4J dataset.

This result highlights the limitations of \system~in fixing specific bugs, which require multiple modifications, requires insertions of statements, are not repetitive in code repositories, and so on.
\textsf{CODIT}, a tree2tree model-based change generation technique, reported several correct modifications related to method arguments, for Defects4J~\cite{DBLP:journals/corr/abs-1810-00314}, but it did not generate fixes in Table~\ref{tab:defects4jresult}. Hence, we believe that our seq2seq approach can be a complement to tree2tree approaches.


\subsection{When/Why Does NMT Fail?}
\label{ssec:challenge}


We see that NMT can work better than the pattern-based patch suggestion for learning past changes and generating fixes. However, we also find limitations of our approach using NMT for code repository data. We examined some of the cases where our approach failed and discuss the challenges (and possible improvements) from two aspects, modeling and training.

\textbf{Modeling:}
Although NMT can learn the semantic and structural information by taking global context into consideration~\cite{He:2016:INM:3015812.3015835}, some limitations of NMT are known and studied~\cite{He:2016:INM:3015812.3015835,AAAI1714451}.

\textit{Out-of-vocabulary problem} or \textit{UNK problem}.
NMT usually uses the top-N most frequent words in the training data and regards other words as unseen ones, \textit{UNK}. Therefore NMT often makes mistakes in translating low-frequency words. While we alleviated this problem somewhat by incorporating dictionaries into our method~\cite{ArthurNN16}, we still find similar issues in low-frequency or novel identifier names, as discussed in the survey result of \textit{challenging fix}. Since those names can be identified from the context, integrating NMT with program analysis could be a promising direction.

\textit{Coverage problem}.
NMT lacks a mechanism to guarantee that all words in a source statement are translated, and usually favors short translations. For example, in translation of long method chains with low-frequency tokens, we see insufficient outputs including incomplete statements and disappearing method calls. Moreover, this problem makes it difficult to address larger fix generation for more than one line. As there are several studies trying to address this problem~\cite{AAAI1714161,DBLP:journals/corr/LiuUFS16,DBLP:journals/corr/WuSCLNMKCGMKSJL16}, we can consider applying these rapidly developed techniques.

\textbf{Training:} 
In addition to techniques related to NMT, we think there is room for improvement in preparing training data.
%
%
In this study, we design the experiment as batch learning, that is, whole training data is prepared from the past until the previous year of the test year. However, Barr \emph{et al.} reported that more reusable pieces of code exist in the immediately previous version~\cite{Barr:2014:PSH:2635868.2635898}. Previous studies have tried an online learning setting called training on errors~\cite{Mizuno:2007:TEE:1287624.1287683,Hata:2008:EFF:1370750.1370772}. Applying such online learning could be a promising challenge too.

\subsection{Limitations and Challenges}


Limitations of \system~are summarized as follows.
\begin{enumerate}
    \item It cannot generate patches that need additions or deletions of statements.
    \item It cannot generate patches that consist of multiple statements.
    \item It cannot generate patches that include the changes of method or array arguments.
    \item It cannot account for the context of patches outside statements.
    \item It cannot generate patches that include unknown tokens.
    \item It cannot generate assembled patches for single bugs.
    \item It is not evaluated in a cross-project setting.
    \item It does not include a bug-localization step.
\end{enumerate}
Considering the amount of hunks with single-statement pairs in Table~\ref{tab:stat} (66.3\% in average), the filtering results in Table~\ref{tab:train} (40.4\% in average), the percentage of NU in Table~\ref{tab:test} (41.0\% in average), around 11\% of change hunks can be targeted in our approach.

Regarding 1), 2), 3), and 4), extending the granularity of code to be learned can address these issues to some extent. As we discuss in Section~\ref{ssec:nmtbased}, some studies reported that learning the contents of methods or classes can work with seq2seq models~\cite{Tufano:2019:LMC:3339505.3339509,DBLP:journals/corr/abs-1901-01808}. More advanced tree-to-tree models may also applicable to address these limitations~\cite{DBLP:journals/corr/abs-1810-00314}.
Regarding 5), increasing appropriate training data is one direction, as well as accounting for context information.
Regarding 6), recent studies have examined multi-hunk program repair~\cite{Saha:2019:HEM:3339505.3339508}. Our fine-grained code history analysis could be suitable for their approach of using revision histories.
Although this paper does not conduct the evaluation of cross-project learning (limitation 7)), \system~can learn from data from multiple projects. Selecting and preprocessing data from multiple projects in an appropriate way can be a practical challenge.

Regarding 8), there are studies of fine-grained defect prediction~\cite{Kim:2008:CSC:1399105.1399451,Hata:2012:BPB:2337223.2337247,Ray:2016:NBC:2884781.2884848}. To localize problematic statements, we could try applying these techniques. We could also consider training another neural network to identify problematic statements to be query statements to \system.

\subsection{Threats to Validity}
Concerning \textit{external validity}, this study only targets five open source projects written in Java.
As we do not have clear selection criteria, there can be a selection bias. Projects with different sizes, different management governance, etc. can lead to different results.
Regarding programming languages, there is a threat of generalization, and it should be interesting to extend this study to other languages.

With respect to \textit{construct validity}, we collect fixes from histories, which can contain mistakes. For example, sometimes fixes can be reverted, but we do not consider such intention.
In addition, the SZZ algorithm used for identifying bug-inducing and bug-fixing commits is known to produce errors~\cite{7588121}. Although we do not distinguish buggy and non-buggy changes for training, we classify test data as buggy or non-buggy. This could impact our discussion regarding the type of changes. However, as presented in Section~\ref{ssec:nonbug}, \system~can work for generation of non-bug-fixing statements as well as bug-fixing statements.

Another threat to \textit{construct validity} exists in our preprocessing.
We removed short statements (step (3)), the contents of method and array arguments (step (4)), unparsable statements (step (5)), and older and less frequent post-correction statements from multiple candidates (step (6)), to ignore noises.
Although these steps were prepared in our trial-and-error experiments and evaluated empirically, different parameters and processes may improve the performance. Exploring a better configuration of preprocessing can be practical future work.

To mitigate threats to \textit{reliability}, we made our dataset and survey material publicly available (see Section~\ref{ssec:data} and Section~\ref{sec:eval}).

\section{Related Work}
\label{sec:related_work}

In this section, we discuss similar NMT-based patch generation studies, and then discuss
two research areas, namely probabilistic models of source code and change mining, which are typically used to build models by learning and mining data for learning.

\subsection{NMT-based Patch Generation}
\label{ssec:nmtbased}

Learning source code changes using NMT-based techniques, NMT-based automated code changes, is an emerging research topic.
Tufano \emph{et al.} conducted empirical studies to investigate the feasibility of learning bug-fixing patches using NMT techniques~\cite{Tufano:2018:EIL:3238147.3240732,2018arXiv181208693T}.
Similar to our approach, they built seq2seq models.
By extending those studies, Tufano \emph{et al.} studied the ability of a seq2seq NMT model to automate code changes for pull requests~\cite{Tufano:2019:LMC:3339505.3339509}.
One of the differences of the earlier approaches and \system~is the granularity of code to be learned.
While the above studies targeted methods within 50 or 100 tokens, \system~targets statements. Positive results at both granularity leveles show the capability of NMT models to learn different types of code changes. Although Tufano \emph{et al.} prepared their training and testing data by random partitioning~\cite{Tufano:2019:LMC:3339505.3339509}, we prepared data considering chronological order, to emulate practical scenarios.
Tufano \emph{et al.} largely abstracted tokens for cross-project learning, while \system~kept identifiers and literals except for arguments to learn project-specific time-sensitive changes, which results in the successful number incrementation shown at the first result in Table~\ref{tab:example}.

\textsf{SequenceR} is another NMT-based system to learn source code changes based on a seq2seq model and copy mechanism~\cite{DBLP:journals/corr/abs-1901-01808}. Similar to \system, one-line changes are targeted for fixing.
While \system~expects only a buggy line as an input, \textsf{SequenceR} accepts surrounding method and class as well as an annotated buggy line as the context.
\textsf{SequenceR} is an end-to-end approach including validation with test cases.
\textsf{CODIT} learns source code change patterns with tree-to-tree NMT models considering AST-level changes~\cite{DBLP:journals/corr/abs-1810-00314}.
Similar to Tufano \emph{et al.}~\cite{Tufano:2019:LMC:3339505.3339509}, cross-project datasets without considering time were used for the evaluation of both \textsf{SequenceR} and \textsf{CODIT}.

While making the vocabulary small is considered to be one of challenges in other studies~\cite{Tufano:2019:LMC:3339505.3339509,DBLP:journals/corr/abs-1901-01808,DBLP:journals/corr/abs-1810-00314}, we did not explicitly limit the number of tokens or identifiers to be learned. There seems to be a trade-off relationship between the vocabulary size and the context size. Since we targeted almost the smallest context (single-statement changes and changes within a single project), we did not need to make the vocabulary small. Addressing this trade-off to consider both larger vocabulary and wider context will be challenging future work.
%
Another major difference between this study and the other studies is our human evaluation with the survey. We observed that sometimes the survey participants perceived positively even if generated statements were not identical to actual statements. Further user studies in practical scenario could be another future challenge.

\begin{table*}[!t]
\renewcommand{\arraystretch}{1.3}
\caption{Studies on source code generating models. The column \textit{Data} is presented by the authors and other contents were previously presented by Allamanis \emph{et al.}~\cite{2017arXiv170906182A}. Only referred papers are presented.}
\label{tab:survey}
\centering
\begin{tabularx}{\linewidth} {lllll}
\toprule
Study & Representation & Model & Application & Data \\
\midrule
Allamanis and Sutton~\cite{Allamanis:2013:MSC:2487085.2487127} & Token & n-gram & --- & Snapshot (Java) \\
Allamanis \emph{et al.}~\cite{Allamanis:2014:LNC:2635868.2635883} & Token + Location & n-gram & Coding conventions & Snapshot (Java) \\
Allamanis and Sutton~\cite{Allamanis:2014:MIS:2635868.2635901} & Syntax & Grammar (pTSG) & --- & Snapshot (Java) \\
Allamanis \emph{et al.}~\cite{Allamanis:2015:BMS:3045118.3045344} & Syntax & Grammar (NN-LBL) & Code search/synthesis & Stack Overflow (C\# and NL) \\
Bielik \emph{et al.}~\cite{Bielik:2016:PPM:3045390.3045699} & Syntax & PCFG + annotations & Code completion & Snapshot (JavaScript) \\
Campbell \emph{et al.}~\cite{Campbell:2014:SEJ:2597073.2597102} & Token & n-gram & Syntax error detection & Selected versions (Java) \\
Cerulo \emph{et al.}~\cite{Cerulo:2015:IRI:2797418.2797556} & Token & Graphical model (HMM) & Information extraction & Snapshot (Java) and NL \\
Cummins \emph{et al.}~\cite{Cummins:2017:SBP:3049832.3049843} & Character & NN (LSTM) & Benchmark synthesis & Benchmarks (OpenCL) \\
Gulwani and Marron~\cite{Gulwani:2014:NIP:2588555.2612177} & Syntax & Phrase model & Text-to-code & Created (DSL and NL) \\
Gvero and Kuncak~\cite{Gvero:2015:SJE:2814270.2814295} & Syntax & PCFG + Search & Code synthesis & Created (Java and NL) \\
Hata \emph{et al.}~\cite{Hata:2008:EFF:1370750.1370772} & Token & Orthogonal sparse bigrams & Bug prediction & Long period (Java) \\
Hata \emph{et al.}~\cite{Hata:2010:FMD:1743628.1743651} & Token & Vector space model & Bug prediction & Snapshot (Java) \\
Hellendoorn \emph{et al.}~\cite{Hellendoorn:2015:TLT:2820518.2820539}  & Token & n-gram & Code review & Short period (Java) \\
Hellendoorn and Devanbu~\cite{Hellendoorn:2017:DNN:3106237.3106290} & Token & n-gram (cache) & --- & Snapshot (Java) \\
Hindle \emph{et al.}~\cite{Hindle:2012:NS:2337223.2337322} & Token & n-gram & Code completion & Snapshot (Java and C) \\
Hsiao \emph{et al.}~\cite{Hsiao:2014:UWC:2660193.2660226} & PDG & n-gram & Program analysis & Snapshot (JavaScript) \\
Ling \emph{et al.}~\cite{ling-EtAl:2016:P16-1} & Token & RNN + attention & Code synthesis & Snapshot (Java and Python) \\ 
Karaivanov \emph{et al.}~\cite{Karaivanov:2014:PST:2661136.2661148} & Token & Phrase & Migration & Snapshot (C\# and Java) \\
Kushman and Barzilay~\cite{kushman-barzilay:2013:NAACL-HLT} & Token & Grammar (CCG) & Code synthesis & Created (Regex and NL) \\
Maddison and Tarlow~\cite{Maddison:2014:SGM:3044805.3044965} & Syntax with scope & NN & --- & TopCoder (C\#) \\
Menon \emph{et al.}~\cite{Menon:2013:MLF:3042817.3042840} & Syntax & PCFG + annotations & Code synthesis & Excel help forums \\
Mizuno and Kikuno~\cite{Mizuno:2007:TEE:1287624.1287683} & Token & Orthogonal sparse bigrams & Bug prediction & Long period (Java) \\
Nguyen \emph{et al.}~\cite{Nguyen:2013:SSL:2491411.2491458} & Token + parse info & n-gram & Code completion & Snapshot (Java) \\
Nguyen \emph{et al.}~\cite{Nguyen:2015:DAM:2916135.2916174} & Token + parse info & Phrase SMT & Migration & Snapshot (Java and C\#) \\
Nguyen and Nguyen~\cite{Nguyen:2015:GSL:2818754.2818858} & Partial PDG & n-gram & Code completion & Snapshot (Java and C\#) \\
Nguyen \emph{et al.}~\cite{Nguyen:2016:LAU:2884781.2884873} & Bytecode & Graphical model (HMM) & Code completion & Android \\
Oda \emph{et al.}~\cite{Oda:2015:LGP:2916135.2916173} & Syntax + token & Tree-to-string + phrase & Pseudocode generation & Created (Python and NL) \\
Rabinovich \emph{et al.}~\cite{rabinovich-stern-klein:2017:Long} & Syntax & NN (LSTM-based) & Code synthesis & Snapshot (Java and Python) \\
Ray \emph{et al.}~\cite{Ray:2016:NBC:2884781.2884848} & Token & n-gram (cache) & Bug detection & Short period (Java) \\
Raychev \emph{et al.}~\cite{Raychev:2014:CCS:2594291.2594321} & Token + constraints & n-gram / RNN & Code completion & Android \\
Raychev \emph{et al.}~\cite{Raychev:2016:LPN:2837614.2837671} & Syntax & PCFG + annotations & Code completion & Snapshot (JavaScript) \\
Sharma \emph{et al.}~\cite{7081855} & Token & n-gram & Information extraction & Stack Overflow and Twitter \\
Tu \emph{et al.}~\cite{Tu:2014:LS:2635868.2635875} & Token & n-gram (cache) & Code completion & Snapshot (Java and Python) \\
Vasilescu \emph{et al.}~\cite{Vasilescu:2017:RCN:3106237.3106289} & Token & Phrase SMT & Deobfuscation & Snapshot (JavaScript) \\
White \emph{et al.}~\cite{White:2015:TDL:2820518.2820559} & Token & NN (RNN) & --- & Snapshot (Java) \\
Yadid and Yahav~\cite{Yadid:2016:ECP:2986012.2986021} & Token & n-gram & Information extraction & Android tutorial videos \\
Yin and Neubig~\cite{yin-neubig:2017:Long} & Syntax & NN (seq2seq) & Synthesis & Snapshot (Python and DSL) \\
\textbf{\system} & Syntax & NN (seq2seq) & Patch generation & Long period (Java) \\
\bottomrule
\end{tabularx}
\end{table*}

\subsection{Probabilistic Models of Source Code}

There are several studies on probabilistic machine learning models of source code for different applications using different techniques. Allamanis \emph{et al.} conducted a large survey on this topic~\cite{2017arXiv170906182A}.
Table~\ref{tab:survey} is originally presented in the survey of representative code models~\cite{2017arXiv170906182A}. From the original table, non-refereed papers are excluded, some missing papers are added, and the column Data is newly prepared, which summarizes analyzed data in terms of programing languages, data sources, and historical information.

As we see from the table, probabilistic machine learning models have been studied for various applications, such as code completion, code synthesis, coding conventions, and so on. From the point of view of models, newer techniques of neural networks (NN), especially neural seq2seq models, have not been extensively studied yet. So there are possibilities of extending and improving previous studies applying these models.

From the data column, we see that several programing languages have been studied including Java, C, C\#, JavaScript, Python, among others. Although most of studies collected data from code repositories, some used other data sources,
for example, programs in TopCorder.com~\cite{Maddison:2014:SGM:3044805.3044965},
Microsoft Excel help forums~\cite{Menon:2013:MLF:3042817.3042840},
Android programming tutorial videos~\cite{Yadid:2016:ECP:2986012.2986021},
to build probabilistic models of source code.
From source code repositories, collecting source code in selected snapshots is a common procedure.
However, when considering software evolution, that is, software is updated continuously, learning over long periods is more practical.
As discussed in Section~\ref{ssec:challenge}, online machine learning is one of challenges in this scenario.
Previous studies demonstrated learning methods in long periods, called training on errors~\cite{Mizuno:2007:TEE:1287624.1287683,Hata:2008:EFF:1370750.1370772}. This can be a good hint for future research on online machine learning of patch generation.

\subsection{Change Mining}

Analyzing and exploiting historical change patterns is another similar topic to this work.
Kim \emph{et al.} proposed bug finding techniques based on textual code change histories~\cite{Kim:2006:MBF:1181775.1181781}. From the analysis of open source repositories, they reported that a large amount of bugs appeared repeatedly.
From the analysis of graph-based object usage models, Nguyen \emph{et al.} also reported recurring bug-fix patterns and demonstrated fix recommendation based on those patterns~\cite{Nguyen:2010:RBF:1806799.1806847}.
To make use of similar code changes, Meng \emph{et al.} proposes a tool called LASE for creating and applying context-aware edit scripts~\cite{Meng:2013:LLA:2486788.2486855}. LASE analyzes AST-level changes and generates AST node edit operations.
From a large-scale study of AST-level code changes in multiple Java projects, Nguyen \emph{et al.} reported that repetitiveness is high for small size changes and similar bug-fix changes repeatedly occurred in cross projects~\cite{Nguyen:2013:SRC:3107656.3107682}.
Barr \emph{et al.} studied the Plastic Surgery hypothesis, that is, same changes already exist in code histories and those changes can be efficiently found and exploited~\cite{Barr:2014:PSH:2635868.2635898}. From line-granular snippet matching analyses, they reported that changes are repetitive and this repetitiveness is usefully exploitable.
Yue \emph{et al.} reported, from an empirical study of large-scale bug fixes, that 15-20\% of bugs involved repeated fixes~\cite{8094441}.

As these studies presented, using change patterns can be promising. However, from the study of the uniqueness of changes, instead of common changes, Ray \emph{et al.} reported that unique changes are more common than non-unique changes~\cite{Ray:2015:UCC:2820518.2820526}. This implies that simply applying past change patterns has limited capabilities in terms of reuse. As our results demonstrated, NMT-based learning approaches have the ability to address this issue by learning bug-fix correspondences on a variety of levels.

\section{Conclusion}
\label{sec:conclusion}
In this paper, we introduced \system, an NMT-based technique to generate bug fixes from past fixes. Through an empirical validation on five open source projects, we find that \system~is effective in generating fixes. Moreover, we show that \system~can even be used to generate statements for non-bug-fixing statements. We compare \system~to pattern-based patch suggestion as a baseline and show that \system~performs at least as well as the baseline.

We also investigate cases where \system~fails and find that \system, or more generally NMT, suffers from the out-of-vocabulary problem since it depends on the presence of words in the past to train on. Also, NMT cannot guarantee that all words are covered/translated. These aforementioned issues are areas that we plan to tackle in future work.


%

\ifCLASSOPTIONcompsoc
  \section*{Acknowledgments}
\else
  \section*{Acknowledgment}
\fi


\ifCLASSOPTIONcaptionsoff
  \newpage
\fi




\begin{thebibliography}{100}
\providecommand{\url}[1]{#1}
\csname url@samestyle\endcsname
\providecommand{\newblock}{\relax}
\providecommand{\bibinfo}[2]{#2}
\providecommand{\BIBentrySTDinterwordspacing}{\spaceskip=0pt\relax}
\providecommand{\BIBentryALTinterwordstretchfactor}{4}
\providecommand{\BIBentryALTinterwordspacing}{\spaceskip=\fontdimen2\font plus
\BIBentryALTinterwordstretchfactor\fontdimen3\font minus
  \fontdimen4\font\relax}
\providecommand{\BIBforeignlanguage}[2]{{%
\expandafter\ifx\csname l@#1\endcsname\relax
\typeout{** WARNING: IEEEtran.bst: No hyphenation pattern has been}%
\typeout{** loaded for the language `#1'. Using the pattern for}%
\typeout{** the default language instead.}%
\else
\language=\csname l@#1\endcsname
\fi
#2}}
\providecommand{\BIBdecl}{\relax}
\BIBdecl

\bibitem{Pei:2014:AFP:2693206.2693287}
\BIBentryALTinterwordspacing
Y.~Pei, C.~A. Furia, M.~Nordio, Y.~Wei, B.~Meyer, and A.~Zeller, ``Automated
  fixing of programs with contracts,'' \emph{{IEEE} Trans. Softw. Eng.},
  vol.~40, no.~5, pp. 427--449, May 2014. [Online]. Available:
  \url{https://doi.org/10.1109/TSE.2014.2312918}
\BIBentrySTDinterwordspacing

\bibitem{LeGoues:2012:GGM:2122269.2122544}
\BIBentryALTinterwordspacing
C.~Le~Goues, T.~Nguyen, S.~Forrest, and W.~Weimer, ``Genprog: A generic method
  for automatic software repair,'' \emph{{IEEE} Trans. Softw. Eng.}, vol.~38,
  no.~1, pp. 54--72, Jan. 2012. [Online]. Available:
  \url{http://dx.doi.org/10.1109/TSE.2011.104}
\BIBentrySTDinterwordspacing

\bibitem{LeGoues:2012:SSA:2337223.2337225}
\BIBentryALTinterwordspacing
C.~Le~Goues, M.~Dewey-Vogt, S.~Forrest, and W.~Weimer, ``A systematic study of
  automated program repair: Fixing 55 out of 105 bugs for \$8 each,'' in
  \emph{Proc.\ of 34th Int. Conf. on Softw. Eng.}, ser. ICSE '12.\hskip 1em
  plus 0.5em minus 0.4em\relax Piscataway, NJ, USA: IEEE Press, 2012, pp.
  3--13. [Online]. Available:
  \url{http://dl.acm.org/citation.cfm?id=2337223.2337225}
\BIBentrySTDinterwordspacing

\bibitem{Liu:2012:AAF:2337223.2337259}
\BIBentryALTinterwordspacing
P.~Liu and C.~Zhang, ``Axis: Automatically fixing atomicity violations through
  solving control constraints,'' in \emph{Proc.\ of 34th Int. Conf. on Softw.
  Eng.}, ser. ICSE '12.\hskip 1em plus 0.5em minus 0.4em\relax Piscataway, NJ,
  USA: IEEE Press, 2012, pp. 299--309. [Online]. Available:
  \url{http://dl.acm.org/citation.cfm?id=2337223.2337259}
\BIBentrySTDinterwordspacing

\bibitem{Nguyen:2013:SPR:2486788.2486890}
\BIBentryALTinterwordspacing
H.~D.~T. Nguyen, D.~Qi, A.~Roychoudhury, and S.~Chandra, ``Semfix: Program
  repair via semantic analysis,'' in \emph{Proc.\ of 35th Int. Conf. on Softw.
  Eng.}, ser. ICSE '13.\hskip 1em plus 0.5em minus 0.4em\relax Piscataway, NJ,
  USA: IEEE Press, 2013, pp. 772--781. [Online]. Available:
  \url{http://dl.acm.org/citation.cfm?id=2486788.2486890}
\BIBentrySTDinterwordspacing

\bibitem{Coker:2013:PTF:2486788.2486892}
\BIBentryALTinterwordspacing
Z.~Coker and M.~Hafiz, ``Program transformations to fix c integers,'' in
  \emph{Proc.\ of 35th Int. Conf. on Softw. Eng.}, ser. ICSE '13.\hskip 1em
  plus 0.5em minus 0.4em\relax Piscataway, NJ, USA: IEEE Press, 2013, pp.
  792--801. [Online]. Available:
  \url{http://dl.acm.org/citation.cfm?id=2486788.2486892}
\BIBentrySTDinterwordspacing

\bibitem{Weimer:2013:LPE:3107656.3107702}
\BIBentryALTinterwordspacing
W.~Weimer, Z.~P. Fry, and S.~Forrest, ``Leveraging program equivalence for
  adaptive program repair: Models and first results,'' in \emph{Proc.\ of 28th
  {IEEE}/{ACM} Int. Conf. on Automated Softw. Eng.}, ser. ASE'13.\hskip 1em
  plus 0.5em minus 0.4em\relax Piscataway, NJ, USA: IEEE Press, 2013, pp.
  356--366. [Online]. Available: \url{https://doi.org/10.1109/ASE.2013.6693094}
\BIBentrySTDinterwordspacing

\bibitem{Qi:2014:SRS:2568225.2568254}
\BIBentryALTinterwordspacing
Y.~Qi, X.~Mao, Y.~Lei, Z.~Dai, and C.~Wang, ``The strength of random search on
  automated program repair,'' in \emph{Proc.\ of 36th Int. Conf. on Softw.
  Eng.}, ser. ICSE 2014.\hskip 1em plus 0.5em minus 0.4em\relax New York, NY,
  USA: ACM, 2014, pp. 254--265. [Online]. Available:
  \url{http://doi.acm.org/10.1145/2568225.2568254}
\BIBentrySTDinterwordspacing

\bibitem{Kaleeswaran:2014:MAS:2568225.2568258}
\BIBentryALTinterwordspacing
S.~Kaleeswaran, V.~Tulsian, A.~Kanade, and A.~Orso, ``Minthint: Automated
  synthesis of repair hints,'' in \emph{Proc.\ of 36th Int. Conf. on Softw.
  Eng.}, ser. ICSE 2014.\hskip 1em plus 0.5em minus 0.4em\relax New York, NY,
  USA: ACM, 2014, pp. 266--276. [Online]. Available:
  \url{http://doi.acm.org/10.1145/2568225.2568258}
\BIBentrySTDinterwordspacing

\bibitem{OcarizaJr.:2014:VSF:2568225.2568257}
\BIBentryALTinterwordspacing
F.~S. Ocariza, Jr., K.~Pattabiraman, and A.~Mesbah, ``Vejovis: Suggesting fixes
  for javascript faults,'' in \emph{Proc.\ of 36th Int. Conf. on Softw. Eng.},
  ser. ICSE 2014.\hskip 1em plus 0.5em minus 0.4em\relax New York, NY, USA:
  ACM, 2014, pp. 837--847. [Online]. Available:
  \url{http://doi.acm.org/10.1145/2568225.2568257}
\BIBentrySTDinterwordspacing

\bibitem{Liu:2014:GCF:2635868.2635881}
\BIBentryALTinterwordspacing
P.~Liu, O.~Tripp, and C.~Zhang, ``Grail: Context-aware fixing of concurrency
  bugs,'' in \emph{Proc.\ of 22nd {ACM} {SIGSOFT} Int. Symp. on Found. of
  Softw. Eng.}, ser. FSE 2014.\hskip 1em plus 0.5em minus 0.4em\relax New York,
  NY, USA: ACM, 2014, pp. 318--329. [Online]. Available:
  \url{http://doi.acm.org/10.1145/2635868.2635881}
\BIBentrySTDinterwordspacing

\bibitem{Lin:2014:ARM:2610384.2610398}
\BIBentryALTinterwordspacing
Y.~Lin and S.~S. Kulkarni, ``Automatic repair for multi-threaded programs with
  deadlock/livelock using maximum satisfiability,'' in \emph{Proc. of 23rd Int.
  Symp. on Softw. Testing and Analysis}, ser. ISSTA 2014.\hskip 1em plus 0.5em
  minus 0.4em\relax New York, NY, USA: ACM, 2014, pp. 237--247. [Online].
  Available: \url{http://doi.acm.org/10.1145/2610384.2610398}
\BIBentrySTDinterwordspacing

\bibitem{Mechtaev:2015:DLS:2818754.2818811}
\BIBentryALTinterwordspacing
S.~Mechtaev, J.~Yi, and A.~Roychoudhury, ``Directfix: Looking for simple
  program repairs,'' in \emph{Proc.\ of 37th Int. Conf. on Softw. Eng. - Vol.
  1}, ser. ICSE '15.\hskip 1em plus 0.5em minus 0.4em\relax Piscataway, NJ,
  USA: IEEE Press, 2015, pp. 448--458. [Online]. Available:
  \url{http://dl.acm.org/citation.cfm?id=2818754.2818811}
\BIBentrySTDinterwordspacing

\bibitem{Gao:2015:SMF:2818754.2818812}
\BIBentryALTinterwordspacing
Q.~Gao, Y.~Xiong, Y.~Mi, L.~Zhang, W.~Yang, Z.~Zhou, B.~Xie, and H.~Mei, ``Safe
  memory-leak fixing for c programs,'' in \emph{Proc.\ of 37th Int. Conf. on
  Softw. Eng. - Vol. 1}, ser. ICSE '15.\hskip 1em plus 0.5em minus 0.4em\relax
  Piscataway, NJ, USA: IEEE Press, 2015, pp. 459--470. [Online]. Available:
  \url{http://dl.acm.org/citation.cfm?id=2818754.2818812}
\BIBentrySTDinterwordspacing

\bibitem{Tan:2015:RAR:2818754.2818813}
\BIBentryALTinterwordspacing
S.~H. Tan and A.~Roychoudhury, ``Relifix: Automated repair of software
  regressions,'' in \emph{Proc.\ of 37th Int. Conf. on Softw. Eng. - Vol. 1},
  ser. ICSE '15.\hskip 1em plus 0.5em minus 0.4em\relax Piscataway, NJ, USA:
  IEEE Press, 2015, pp. 471--482. [Online]. Available:
  \url{http://dl.acm.org/citation.cfm?id=2818754.2818813}
\BIBentrySTDinterwordspacing

\bibitem{Long:2015:SPR:2786805.2786811}
\BIBentryALTinterwordspacing
F.~Long and M.~Rinard, ``Staged program repair with condition synthesis,'' in
  \emph{Proc.\ of 10th Joint Meeting of the European Softw. Eng. Conf. and the
  {ACM} {SIGSOFT} Symp. on the Found. of Softw. Eng.}, ser. ESEC/FSE
  2015.\hskip 1em plus 0.5em minus 0.4em\relax New York, NY, USA: ACM, 2015,
  pp. 166--178. [Online]. Available:
  \url{http://doi.acm.org/10.1145/2786805.2786811}
\BIBentrySTDinterwordspacing

\bibitem{Mechtaev:2016:ASM:2884781.2884807}
\BIBentryALTinterwordspacing
S.~Mechtaev, J.~Yi, and A.~Roychoudhury, ``Angelix: Scalable multiline program
  patch synthesis via symbolic analysis,'' in \emph{Proc.\ of 38th Int. Conf.
  on Softw. Eng.}, ser. ICSE '16.\hskip 1em plus 0.5em minus 0.4em\relax New
  York, NY, USA: ACM, 2016, pp. 691--701. [Online]. Available:
  \url{http://doi.acm.org/10.1145/2884781.2884807}
\BIBentrySTDinterwordspacing

\bibitem{Xuan:2017:NAR:3071893.3071964}
\BIBentryALTinterwordspacing
J.~Xuan, M.~Martinez, F.~DeMarco, M.~Clement, S.~L. Marcote, T.~Durieux,
  D.~Le~Berre, and M.~Monperrus, ``Nopol: Automatic repair of conditional
  statement bugs in java programs,'' \emph{{IEEE} Trans. Softw. Eng.}, vol.~43,
  no.~1, pp. 34--55, Jan. 2017. [Online]. Available:
  \url{https://doi.org/10.1109/TSE.2016.2560811}
\BIBentrySTDinterwordspacing

\bibitem{Xiong:2017:PCS:3097368.3097418}
\BIBentryALTinterwordspacing
Y.~Xiong, J.~Wang, R.~Yan, J.~Zhang, S.~Han, G.~Huang, and L.~Zhang, ``Precise
  condition synthesis for program repair,'' in \emph{Proc.\ of 39th Int. Conf.
  on Softw. Eng.}, ser. ICSE '17.\hskip 1em plus 0.5em minus 0.4em\relax
  Piscataway, NJ, USA: IEEE Press, 2017, pp. 416--426. [Online]. Available:
  \url{https://doi.org/10.1109/ICSE.2017.45}
\BIBentrySTDinterwordspacing

\bibitem{Le:2017:SSS:3106237.3106309}
\BIBentryALTinterwordspacing
X.-B.~D. Le, D.-H. Chu, D.~Lo, C.~Le~Goues, and W.~Visser, ``S3: Syntax- and
  semantic-guided repair synthesis via programming by examples,'' in
  \emph{Proc.\ of 11th Joint Meeting on Found. of Softw. Eng.}, ser. ESEC/FSE
  2017.\hskip 1em plus 0.5em minus 0.4em\relax New York, NY, USA: ACM, 2017,
  pp. 593--604. [Online]. Available:
  \url{http://doi.acm.org/10.1145/3106237.3106309}
\BIBentrySTDinterwordspacing

\bibitem{Saha:2017:EEO:3155562.3155643}
\BIBentryALTinterwordspacing
R.~K. Saha, Y.~Lyu, H.~Yoshida, and M.~R. Prasad, ``Elixir: Effective object
  oriented program repair,'' in \emph{Proc.\ of 32nd {IEEE}/{ACM} Int. Conf. on
  Automated Softw. Eng.}, ser. ASE 2017.\hskip 1em plus 0.5em minus 0.4em\relax
  Piscataway, NJ, USA: IEEE Press, 2017, pp. 648--659. [Online]. Available:
  \url{http://dl.acm.org/citation.cfm?id=3155562.3155643}
\BIBentrySTDinterwordspacing

\bibitem{Hua:2018:TPP:3180155.3180245}
\BIBentryALTinterwordspacing
J.~Hua, M.~Zhang, K.~Wang, and S.~Khurshid, ``Towards practical program repair
  with on-demand candidate generation,'' in \emph{Proc.\ of 40th Int. Conf. on
  Softw. Eng.}, ser. ICSE '18.\hskip 1em plus 0.5em minus 0.4em\relax New York,
  NY, USA: ACM, 2018, pp. 12--23. [Online]. Available:
  \url{http://doi.acm.org/10.1145/3180155.3180245}
\BIBentrySTDinterwordspacing

\bibitem{Mechtaev:2018:SPR:3180155.3180247}
\BIBentryALTinterwordspacing
S.~Mechtaev, M.-D. Nguyen, Y.~Noller, L.~Grunske, and A.~Roychoudhury,
  ``Semantic program repair using a reference implementation,'' in \emph{Proc.\
  of 40th Int. Conf. on Softw. Eng.}, ser. ICSE '18.\hskip 1em plus 0.5em minus
  0.4em\relax New York, NY, USA: ACM, 2018, pp. 129--139. [Online]. Available:
  \url{http://doi.acm.org/10.1145/3180155.3180247}
\BIBentrySTDinterwordspacing

\bibitem{Monperrus:2018:ASR:3177787.3105906}
\BIBentryALTinterwordspacing
M.~Monperrus, ``Automatic software repair: A bibliography,'' \emph{ACM Comput.
  Surv.}, vol.~51, no.~1, pp. 17:1--17:24, Jan. 2018. [Online]. Available:
  \url{http://doi.acm.org/10.1145/3105906}
\BIBentrySTDinterwordspacing

\bibitem{8089448}
L.~Gazzola, D.~Micucci, and L.~Mariani, ``Automatic software repair: A
  survey,'' \emph{{IEEE} Trans. Softw. Eng.}, pp. 1--1, 2018.

\bibitem{safix}
\BIBentryALTinterwordspacing
Y.~Jia, K.~Mao, and M.~Harman. (2018) Finding and fixing software bugs
  automatically with sapfix and sapienz. [Online]. Available:
  \url{https://code.fb.com/developer-tools/finding-and-fixing-software-bugs-automatically-with-sapfix-and
  -sapienz/}
\BIBentrySTDinterwordspacing

\bibitem{Smith:2015:CWD:2786805.2786825}
\BIBentryALTinterwordspacing
E.~K. Smith, E.~T. Barr, C.~Le~Goues, and Y.~Brun, ``Is the cure worse than the
  disease? overfitting in automated program repair,'' in \emph{Proc.\ of 10th
  Joint Meeting of the European Softw. Eng. Conf. and the {ACM} {SIGSOFT} Symp.
  on the Found. of Softw. Eng.}, ser. ESEC/FSE 2015.\hskip 1em plus 0.5em minus
  0.4em\relax New York, NY, USA: ACM, 2015, pp. 532--543. [Online]. Available:
  \url{http://doi.acm.org/10.1145/2786805.2786825}
\BIBentrySTDinterwordspacing

\bibitem{Long:2016:ASS:2884781.2884872}
\BIBentryALTinterwordspacing
F.~Long and M.~Rinard, ``An analysis of the search spaces for generate and
  validate patch generation systems,'' in \emph{Proc.\ of 38th Int. Conf. on
  Softw. Eng.}, ser. ICSE '16.\hskip 1em plus 0.5em minus 0.4em\relax New York,
  NY, USA: ACM, 2016, pp. 702--713. [Online]. Available:
  \url{http://doi.acm.org/10.1145/2884781.2884872}
\BIBentrySTDinterwordspacing

\bibitem{Wen:2018:CPG:3180155.3180233}
\BIBentryALTinterwordspacing
M.~Wen, J.~Chen, R.~Wu, D.~Hao, and S.-C. Cheung, ``Context-aware patch
  generation for better automated program repair,'' in \emph{Proc.\ of 40th
  Int. Conf. on Softw. Eng.}, ser. ICSE '18.\hskip 1em plus 0.5em minus
  0.4em\relax New York, NY, USA: ACM, 2018, pp. 1--11. [Online]. Available:
  \url{http://doi.acm.org/10.1145/3180155.3180233}
\BIBentrySTDinterwordspacing

\bibitem{Kim:2013:APG:2486788.2486893}
\BIBentryALTinterwordspacing
D.~Kim, J.~Nam, J.~Song, and S.~Kim, ``Automatic patch generation learned from
  human-written patches,'' in \emph{Proc.\ of 35th Int. Conf. on Softw. Eng.},
  ser. ICSE '13.\hskip 1em plus 0.5em minus 0.4em\relax Piscataway, NJ, USA:
  IEEE Press, 2013, pp. 802--811. [Online]. Available:
  \url{http://dl.acm.org/citation.cfm?id=2486788.2486893}
\BIBentrySTDinterwordspacing

\bibitem{Martinez:2015:MSR:2727049.2727080}
\BIBentryALTinterwordspacing
M.~Martinez and M.~Monperrus, ``Mining software repair models for reasoning on
  the search space of automated program fixing,'' \emph{Empirical Softw. Eng.},
  vol.~20, no.~1, pp. 176--205, Feb. 2015. [Online]. Available:
  \url{http://dx.doi.org/10.1007/s10664-013-9282-8}
\BIBentrySTDinterwordspacing

\bibitem{Ke:2015:RPS:2916135.2916260}
\BIBentryALTinterwordspacing
Y.~Ke, K.~T. Stolee, C.~L. Goues, and Y.~Brun, ``Repairing programs with
  semantic code search (t),'' in \emph{Proc.\ of 30th {IEEE}/{ACM} Int. Conf.
  on Automated Softw. Eng.}, ser. ASE '15.\hskip 1em plus 0.5em minus
  0.4em\relax Washington, DC, USA: IEEE Computer Society, 2015, pp. 295--306.
  [Online]. Available: \url{https://doi.org/10.1109/ASE.2015.60}
\BIBentrySTDinterwordspacing

\bibitem{7476644}
X.~B.~D. Le, D.~Lo, and C.~L. Goues, ``History driven program repair,'' in
  \emph{Proc.\ of 23rd {IEEE} Int. Conf. on Softw. Analysis, Evolution and
  Reengineering}, vol.~1, March 2016, pp. 213--224.

\bibitem{Campos:2017:CBP:3200492.3200554}
\BIBentryALTinterwordspacing
E.~C. Campos and M.~A. Maia, ``Common bug-fix patterns: A large-scale
  observational study,'' in \emph{Proc.\ of 11th {ACM}/{IEEE} Int. Symp. on
  Empirical Softw. Eng. and Measurement}, ser. ESEM '17.\hskip 1em plus 0.5em
  minus 0.4em\relax Piscataway, NJ, USA: IEEE Press, 2017, pp. 404--413.
  [Online]. Available: \url{https://doi.org/10.1109/ESEM.2017.55}
\BIBentrySTDinterwordspacing

\bibitem{Long:2017:AIC:3106237.3106253}
\BIBentryALTinterwordspacing
F.~Long, P.~Amidon, and M.~Rinard, ``Automatic inference of code transforms for
  patch generation,'' in \emph{Proc.\ of 11th Joint Meeting on Found. of Softw.
  Eng.}, ser. ESEC/FSE 2017.\hskip 1em plus 0.5em minus 0.4em\relax New York,
  NY, USA: ACM, 2017, pp. 727--739. [Online]. Available:
  \url{http://doi.acm.org/10.1145/3106237.3106253}
\BIBentrySTDinterwordspacing

\bibitem{Xin:2017:LSC:3155562.3155644}
\BIBentryALTinterwordspacing
Q.~Xin and S.~P. Reiss, ``Leveraging syntax-related code for automated program
  repair,'' in \emph{Proc.\ of 32nd {IEEE}/{ACM} Int. Conf. on Automated Softw.
  Eng.}, ser. ASE 2017.\hskip 1em plus 0.5em minus 0.4em\relax Piscataway, NJ,
  USA: IEEE Press, 2017, pp. 660--670. [Online]. Available:
  \url{http://dl.acm.org/citation.cfm?id=3155562.3155644}
\BIBentrySTDinterwordspacing

\bibitem{Jiang:2018:SPR:3213846.3213871}
\BIBentryALTinterwordspacing
J.~Jiang, Y.~Xiong, H.~Zhang, Q.~Gao, and X.~Chen, ``Shaping program repair
  space with existing patches and similar code,'' in \emph{Proc. of 27th {ACM}
  {SIGSOFT} Int. Symp. on Softw. Testing and Analysis}, ser. ISSTA 2018.\hskip
  1em plus 0.5em minus 0.4em\relax New York, NY, USA: ACM, 2018, pp. 298--309.
  [Online]. Available: \url{http://doi.acm.org/10.1145/3213846.3213871}
\BIBentrySTDinterwordspacing

\bibitem{chan2016listen}
W.~Chan, N.~Jaitly, Q.~Le, and O.~Vinyals, ``Listen, attend and spell: A neural
  network for large vocabulary conversational speech recognition,'' in
  \emph{Proceedings of the International Conference on Acoustics, Speech, and
  Signal Processing (ICASSP)}.\hskip 1em plus 0.5em minus 0.4em\relax IEEE,
  2016, pp. 4960--4964.

\bibitem{vinyals2015grammar}
O.~Vinyals, {\L}.~Kaiser, T.~Koo, S.~Petrov, I.~Sutskever, and G.~Hinton,
  ``Grammar as a foreign language,'' in \emph{Proceedings of the 29th Annual
  Conference on Neural Information Processing Systems (NIPS)}, 2015, pp.
  2773--2781.

\bibitem{rush2015neuralattention}
\BIBentryALTinterwordspacing
A.~M. Rush, S.~Chopra, and J.~Weston, ``A neural attention model for
  abstractive sentence summarization,'' in \emph{Proceedings of the 2015
  Conference on Empirical Methods in Natural Language Processing (EMNLP)},
  2015, pp. 379--389. [Online]. Available:
  \url{http://aclweb.org/anthology/D15-1044}
\BIBentrySTDinterwordspacing

\bibitem{iyer16summarizing}
S.~Iyer, I.~Konstas, A.~Cheung, and L.~Zettlemoyer, ``Summarizing source code
  using a neural attention model,'' in \emph{Proceedings of the 54th Annual
  Meeting of the Association for Computational Linguistics (ACL)}, 2016, pp.
  2073--2083.

\bibitem{ling-EtAl:2016:P16-1}
\BIBentryALTinterwordspacing
W.~Ling, P.~Blunsom, E.~Grefenstette, K.~M. Hermann, T.~Ko\v{c}isk\'{y},
  F.~Wang, and A.~Senior, ``Latent predictor networks for code generation,'' in
  \emph{Proc. of 54th Annual Meeting of Association for Computational
  Linguistics (Volume 1: Long Papers)}.\hskip 1em plus 0.5em minus 0.4em\relax
  Berlin, Germany: Association for Computational Linguistics, August 2016, pp.
  599--609. [Online]. Available: \url{http://www.aclweb.org/anthology/P16-1057}
\BIBentrySTDinterwordspacing

\bibitem{yin-neubig:2017:Long}
\BIBentryALTinterwordspacing
P.~Yin and G.~Neubig, ``A syntactic neural model for general-purpose code
  generation,'' in \emph{Proc. of 55th Annual Meeting of Association for
  Computational Linguistics (Volume 1: Long Papers)}.\hskip 1em plus 0.5em
  minus 0.4em\relax Vancouver, Canada: Association for Computational
  Linguistics, July 2017, pp. 440--450. [Online]. Available:
  \url{http://aclweb.org/anthology/P17-1041}
\BIBentrySTDinterwordspacing

\bibitem{rabinovich-stern-klein:2017:Long}
\BIBentryALTinterwordspacing
M.~Rabinovich, M.~Stern, and D.~Klein, ``Abstract syntax networks for code
  generation and semantic parsing,'' in \emph{Proc. of 55th Annual Meeting of
  Association for Computational Linguistics (Volume 1: Long Papers)}.\hskip 1em
  plus 0.5em minus 0.4em\relax Vancouver, Canada: Association for Computational
  Linguistics, July 2017, pp. 1139--1149. [Online]. Available:
  \url{http://aclweb.org/anthology/P17-1105}
\BIBentrySTDinterwordspacing

\bibitem{Gu:2016:DAL:2950290.2950334}
\BIBentryALTinterwordspacing
X.~Gu, H.~Zhang, D.~Zhang, and S.~Kim, ``Deep api learning,'' in \emph{Proc.\
  of 24th {ACM} {SIGSOFT} Int. Symp. on Found. of Softw. Eng.}, ser. FSE
  2016.\hskip 1em plus 0.5em minus 0.4em\relax New York, NY, USA: ACM, 2016,
  pp. 631--642. [Online]. Available:
  \url{http://doi.acm.org/10.1145/2950290.2950334}
\BIBentrySTDinterwordspacing

\bibitem{bhatia2016automated}
S.~Bhatia and R.~Singh, ``Automated correction for syntax errors in programming
  assignments using recurrent neural networks,'' in \emph{Proc. of 2nd Indian
  Workshop on Machine Learning}, 2016.

\bibitem{AAAI1714603}
\BIBentryALTinterwordspacing
R.~Gupta, S.~Pal, A.~Kanade, and S.~Shevade, ``Deepfix: Fixing common c
  language errors by deep learning,'' in \emph{AAAI Conference on Artificial
  Intelligence}, 2017. [Online]. Available:
  \url{https://aaai.org/ocs/index.php/AAAI/AAAI17/paper/view/14603}
\BIBentrySTDinterwordspacing

\bibitem{koehn10smt}
P.~Koehn, \emph{Statistical Machine Translation}.\hskip 1em plus 0.5em minus
  0.4em\relax Cambridge Press, 2010.

\bibitem{nguyen2013lexical}
A.~T. Nguyen, T.~T. Nguyen, and T.~N. Nguyen, ``Lexical statistical machine
  translation for language migration,'' in \emph{Proceedings of the 2013 9th
  Joint Meeting on Foundations of Software Engineering}.\hskip 1em plus 0.5em
  minus 0.4em\relax ACM, 2013, pp. 651--654.

\bibitem{Oda:2015:LGP:2916135.2916173}
\BIBentryALTinterwordspacing
Y.~Oda, H.~Fudaba, G.~Neubig, H.~Hata, S.~Sakti, T.~Toda, and S.~Nakamura,
  ``Learning to generate pseudo-code from source code using statistical machine
  translation (t),'' in \emph{Proc.\ of 30th {IEEE}/{ACM} Int. Conf. on
  Automated Softw. Eng.}, ser. ASE '15.\hskip 1em plus 0.5em minus 0.4em\relax
  Washington, DC, USA: IEEE Computer Society, 2015, pp. 574--584. [Online].
  Available: \url{http://dx.doi.org/10.1109/ASE.2015.36}
\BIBentrySTDinterwordspacing

\bibitem{Barr:2014:PSH:2635868.2635898}
\BIBentryALTinterwordspacing
E.~T. Barr, Y.~Brun, P.~Devanbu, M.~Harman, and F.~Sarro, ``The plastic surgery
  hypothesis,'' in \emph{Proc.\ of 22nd {ACM} {SIGSOFT} Int. Symp. on Found. of
  Softw. Eng.}, ser. FSE 2014.\hskip 1em plus 0.5em minus 0.4em\relax New York,
  NY, USA: ACM, 2014, pp. 306--317. [Online]. Available:
  \url{http://doi.acm.org/10.1145/2635868.2635898}
\BIBentrySTDinterwordspacing

\bibitem{Rolim:2017:LSP:3097368.3097417}
\BIBentryALTinterwordspacing
R.~Rolim, G.~Soares, L.~D'Antoni, O.~Polozov, S.~Gulwani, R.~Gheyi, R.~Suzuki,
  and B.~Hartmann, ``Learning syntactic program transformations from
  examples,'' in \emph{Proc.\ of 39th Int. Conf. on Softw. Eng.}, ser. ICSE
  '17.\hskip 1em plus 0.5em minus 0.4em\relax Piscataway, NJ, USA: IEEE Press,
  2017, pp. 404--415. [Online]. Available:
  \url{https://doi.org/10.1109/ICSE.2017.44}
\BIBentrySTDinterwordspacing

\bibitem{vanTonder:2018:SAP:3180155.3180250}
\BIBentryALTinterwordspacing
R.~van Tonder and C.~L. Goues, ``Static automated program repair for heap
  properties,'' in \emph{Proc.\ of 40th Int. Conf. on Softw. Eng.}, ser. ICSE
  '18.\hskip 1em plus 0.5em minus 0.4em\relax New York, NY, USA: ACM, 2018, pp.
  151--162. [Online]. Available:
  \url{http://doi.acm.org/10.1145/3180155.3180250}
\BIBentrySTDinterwordspacing

\bibitem{8565907}
K.~{Liu}, D.~{Kim}, T.~F. {Bissyande}, S.~{Yoo}, and Y.~{Le Traon}, ``Mining
  fix patterns for findbugs violations,'' \emph{IEEE Transactions on Software
  Engineering}, pp. 1--1, 2018.

\bibitem{Motwani2018}
\BIBentryALTinterwordspacing
M.~Motwani, S.~Sankaranarayanan, R.~Just, and Y.~Brun, ``Do automated program
  repair techniques repair hard and important bugs?'' \emph{Empirical Softw.
  Eng.}, vol.~23, no.~5, pp. 2901--2947, Oct 2018. [Online]. Available:
  \url{https://doi.org/10.1007/s10664-017-9550-0}
\BIBentrySTDinterwordspacing

\bibitem{Ray:2015:UCC:2820518.2820526}
\BIBentryALTinterwordspacing
B.~Ray, M.~Nagappan, C.~Bird, N.~Nagappan, and T.~Zimmermann, ``The uniqueness
  of changes: Characteristics and applications,'' in \emph{Proc.\ of 12th Work.
  Conf. on Mining Softw. Repositories}, ser. MSR '15.\hskip 1em plus 0.5em
  minus 0.4em\relax Piscataway, NJ, USA: IEEE Press, 2015, pp. 34--44.
  [Online]. Available: \url{http://dl.acm.org/citation.cfm?id=2820518.2820526}
\BIBentrySTDinterwordspacing

\bibitem{kalchbrenner13rnntm}
N.~Kalchbrenner and P.~Blunsom, ``Recurrent continuous translation models,'' in
  \emph{Proceedings of the 2013 Conference on Empirical Methods in Natural
  Language Processing (EMNLP)}, 2013, pp. 1700--1709.

\bibitem{sutskever14sequencetosequence}
I.~Sutskever, O.~Vinyals, and Q.~V. Le, ``Sequence to sequence learning with
  neural networks,'' in \emph{Proceedings of the 28th Annual Conference on
  Neural Information Processing Systems (NIPS)}, 2014, pp. 3104--3112.

\bibitem{neubig2017neural}
G.~Neubig, ``Neural machine translation and sequence-to-sequence models: A
  tutorial,'' \emph{arXiv preprint arXiv:1703.01619}, 2017.

\bibitem{goodfellow2016deep}
I.~Goodfellow, Y.~Bengio, and A.~Courville, \emph{Deep learning}.\hskip 1em
  plus 0.5em minus 0.4em\relax MIT press, 2016.

\bibitem{hornik1989multilayer}
K.~Hornik, M.~Stinchcombe, and H.~White, ``Multilayer feedforward networks are
  universal approximators,'' \emph{Neural networks}, vol.~2, no.~5, pp.
  359--366, 1989.

\bibitem{rumelhart1988learning}
D.~E. Rumelhart, G.~E. Hinton, and R.~J. Williams, ``Learning representations
  by back-propagating errors,'' \emph{Cognitive modeling}, vol.~5, no.~3, p.~1,
  1988.

\bibitem{abadi2016tensorflow}
M.~Abadi, A.~Agarwal, P.~Barham, E.~Brevdo, Z.~Chen, C.~Citro, G.~S. Corrado,
  A.~Davis, J.~Dean, M.~Devin \emph{et~al.}, ``Tensorflow: Large-scale machine
  learning on heterogeneous distributed systems,'' \emph{arXiv preprint
  arXiv:1603.04467}, 2016.

\bibitem{neubig17dynet}
G.~Neubig, C.~Dyer, Y.~Goldberg, A.~Matthews, W.~Ammar, A.~Anastasopoulos,
  M.~Ballesteros, D.~Chiang, D.~Clothiaux, T.~Cohn, K.~Duh, M.~Faruqui, C.~Gan,
  D.~Garrette, Y.~Ji, L.~Kong, A.~Kuncoro, G.~Kumar, C.~Malaviya, P.~Michel,
  Y.~Oda, M.~Richardson, N.~Saphra, S.~Swayamdipta, and P.~Yin, ``Dynet: The
  dynamic neural network toolkit,'' \emph{arXiv preprint arXiv:1701.03980},
  2017.

\bibitem{bahdanau15alignandtranslate}
D.~Bahdanau, K.~Cho, and Y.~Bengio, ``Neural machine translation by jointly
  learning to align and translate,'' in \emph{Proceedings of the International
  Conference on Learning Representations (ICLR)}, 2015.

\bibitem{hochreiter97lstm}
S.~Hochreiter and J.~Schmidhuber, ``Long short-term memory,'' \emph{Neural
  computation}, vol.~9, no.~8, pp. 1735--1780, 1997.

\bibitem{socher11dynamic}
R.~Socher, E.~H. Huang, J.~Pennin, C.~D. Manning, and A.~Y. Ng, ``Dynamic
  pooling and unfolding recursive autoencoders for paraphrase detection,'' in
  \emph{Proceedings of the 25th Annual Conference on Neural Information
  Processing Systems (NIPS)}, 2011, pp. 801--809.

\bibitem{huang2013learning}
P.-S. Huang, X.~He, J.~Gao, L.~Deng, A.~Acero, and L.~Heck, ``Learning deep
  structured semantic models for web search using clickthrough data,'' in
  \emph{Proceedings of the 22nd ACM international conference on Conference on
  information \& knowledge management}.\hskip 1em plus 0.5em minus 0.4em\relax
  ACM, 2013, pp. 2333--2338.

\bibitem{luong15effectiveattentional}
T.~Luong, H.~Pham, and C.~D. Manning, ``Effective approaches to attention-based
  neural machine translation,'' in \emph{Proceedings of the 2015 Conference on
  Empirical Methods in Natural Language Processing (EMNLP)}, 2015, pp.
  1412--1421.

\bibitem{gu-EtAl:2016:P16-1}
\BIBentryALTinterwordspacing
J.~Gu, Z.~Lu, H.~Li, and V.~O. Li, ``Incorporating copying mechanism in
  sequence-to-sequence learning,'' in \emph{Proceedings of the 54th Annual
  Meeting of the Association for Computational Linguistics (Volume 1: Long
  Papers)}.\hskip 1em plus 0.5em minus 0.4em\relax Berlin, Germany: Association
  for Computational Linguistics, August 2016, pp. 1631--1640. [Online].
  Available: \url{http://www.aclweb.org/anthology/P16-1154}
\BIBentrySTDinterwordspacing

\bibitem{arthur16emnlp}
P.~Arthur, G.~Neubig, and S.~Nakamura, ``Incorporating discrete translation
  lexicons into neural machine translation,'' in \emph{Proceedings of the 2016
  Conference on Empirical Methods in Natural Language Processing (EMNLP)},
  2016.

\bibitem{dyer2013simple}
C.~Dyer, V.~Chahuneau, and N.~A. Smith, ``A simple, fast, and effective
  reparameterization of ibm model 2,'' in \emph{Proceedings of the 2013
  Conference of the North American Chapter of the Association for Computational
  Linguistics: Human Language Technologies}, 2013, pp. 644--648.

\bibitem{neubig15lamtram}
G.~Neubig, ``lamtram: A toolkit for language and translation modeling using
  neural networks,'' http://www.github.com/neubig/lamtram, 2015.

\bibitem{kingma2014adam}
D.~Kingma and J.~Ba, ``Adam: A method for stochastic optimization,''
  \emph{arXiv preprint arXiv:1412.6980}, 2014.

\bibitem{srivastava2014dropout}
N.~Srivastava, G.~E. Hinton, A.~Krizhevsky, I.~Sutskever, and R.~Salakhutdinov,
  ``Dropout: a simple way to prevent neural networks from overfitting.''
  \emph{Journal of machine learning research}, vol.~15, no.~1, pp. 1929--1958,
  2014.

\bibitem{Hata:2011:HFV:2024445.2024463}
\BIBentryALTinterwordspacing
H.~Hata, O.~Mizuno, and T.~Kikuno, ``Historage: Fine-grained version control
  system for {Java},'' in \emph{Proc.\ of 3rd Joint Int. and Annual {ERCIM}
  Workshops on Principles of Softw. Evolution and Softw. Evolution Workshops},
  ser. IWPSE-EVOL '11.\hskip 1em plus 0.5em minus 0.4em\relax New York, NY,
  USA: ACM, 2011, pp. 96--100. [Online]. Available:
  \url{http://doi.acm.org/10.1145/2024445.2024463}
\BIBentrySTDinterwordspacing

\bibitem{Fujiwara:2014:KHS:2597073.2597125}
\BIBentryALTinterwordspacing
K.~Fujiwara, H.~Hata, E.~Makihara, Y.~Fujihara, N.~Nakayama, H.~Iida, and
  K.~Matsumoto, ``Kataribe: A hosting service of historage repositories,'' in
  \emph{Proc.\ of 11th Work. Conf. on Mining Softw. Repositories}, ser. MSR
  '14.\hskip 1em plus 0.5em minus 0.4em\relax New York, NY, USA: ACM, 2014, pp.
  380--383. [Online]. Available:
  \url{http://doi.acm.org/10.1145/2597073.2597125}
\BIBentrySTDinterwordspacing

\bibitem{7550864}
K.~Uemura, Y.~Saito, S.~Fujiwara, D.~Tanaka, K.~Fujiwara, H.~Iida, and
  K.~Matsumoto, ``A hosting service of multi-language historage repositories,''
  in \emph{Proc. of {IEEE}/{ACIS} 15th Int. Conf. on Comput. and Inf. Sci.
  (ICIS 2016)}, June 2016, pp. 1--6.

\bibitem{2019arXiv190202467S}
Y.~{Sulistyo Nugroho}, H.~{Hata}, and K.~{Matsumoto}, ``{How Different Are
  Different diff Algorithms in Git? Use --histogram for Code Changes},''
  \emph{arXiv e-prints}, p. arXiv:1902.02467, Feb 2019.

\bibitem{Nguyen:2013:SRC:3107656.3107682}
\BIBentryALTinterwordspacing
H.~A. Nguyen, A.~T. Nguyen, T.~T. Nguyen, T.~N. Nguyen, and H.~Rajan, ``A study
  of repetitiveness of code changes in software evolution,'' in \emph{Proc.\ of
  28th {IEEE}/{ACM} Int. Conf. on Automated Softw. Eng.}, ser. ASE'13.\hskip
  1em plus 0.5em minus 0.4em\relax Piscataway, NJ, USA: IEEE Press, 2013, pp.
  180--190. [Online]. Available: \url{https://doi.org/10.1109/ASE.2013.6693078}
\BIBentrySTDinterwordspacing

\bibitem{4556130}
A.~{Alali}, H.~{Kagdi}, and J.~I. {Maletic}, ``What's a typical commit? a
  characterization of open source software repositories,'' in \emph{2008 16th
  IEEE International Conference on Program Comprehension}, June 2008, pp.
  182--191.

\bibitem{4755633}
O.~{Arafat} and D.~{Riehle}, ``The commit size distribution of open source
  software,'' in \emph{2009 42nd Hawaii International Conference on System
  Sciences}, Jan 2009, pp. 1--8.

\bibitem{sennrich-etal-2016-neural}
\BIBentryALTinterwordspacing
R.~Sennrich, B.~Haddow, and A.~Birch, ``Neural machine translation of rare
  words with subword units,'' in \emph{Proceedings of the 54th Annual Meeting
  of the Association for Computational Linguistics (Volume 1: Long
  Papers)}.\hskip 1em plus 0.5em minus 0.4em\relax Berlin, Germany: Association
  for Computational Linguistics, Aug. 2016, pp. 1715--1725. [Online].
  Available: \url{https://www.aclweb.org/anthology/P16-1162}
\BIBentrySTDinterwordspacing

\bibitem{Rosen:2015:CGA:2786805.2803183}
\BIBentryALTinterwordspacing
C.~Rosen, B.~Grawi, and E.~Shihab, ``Commit guru: Analytics and risk prediction
  of software commits,'' in \emph{Proc.\ of 10th Joint Meeting of the European
  Softw. Eng. Conf. and the {ACM} {SIGSOFT} Symp. on the Found. of Softw.
  Eng.}, ser. ESEC/FSE 2015.\hskip 1em plus 0.5em minus 0.4em\relax New York,
  NY, USA: ACM, 2015, pp. 966--969. [Online]. Available:
  \url{http://doi.acm.org/10.1145/2786805.2803183}
\BIBentrySTDinterwordspacing

\bibitem{Sliwerski:2005:CIF:1083142.1083147}
\BIBentryALTinterwordspacing
J.~\'{S}liwerski, T.~Zimmermann, and A.~Zeller, ``When do changes induce
  fixes?'' in \emph{Proc.\ of 2nd Int. Workshop on Mining Softw. Repositories},
  ser. MSR '05.\hskip 1em plus 0.5em minus 0.4em\relax New York, NY, USA: ACM,
  2005, pp. 1--5. [Online]. Available:
  \url{http://doi.acm.org/10.1145/1082983.1083147}
\BIBentrySTDinterwordspacing

\bibitem{Siegmund:2014:MMP:2674501.2674544}
\BIBentryALTinterwordspacing
J.~Siegmund, C.~K\"{a}stner, J.~Liebig, S.~Apel, and S.~Hanenberg, ``Measuring
  and modeling programming experience,'' \emph{Empirical Softw. Engg.},
  vol.~19, no.~5, pp. 1299--1334, Oct. 2014. [Online]. Available:
  \url{http://dx.doi.org/10.1007/s10664-013-9286-4}
\BIBentrySTDinterwordspacing

\bibitem{Just:2014:DDE:2610384.2628055}
\BIBentryALTinterwordspacing
R.~Just, D.~Jalali, and M.~D. Ernst, ``Defects4j: A database of existing faults
  to enable controlled testing studies for java programs,'' in \emph{Proc. of
  23rd Int. Symp. on Softw. Testing and Analysis}, ser. ISSTA 2014.\hskip 1em
  plus 0.5em minus 0.4em\relax New York, NY, USA: ACM, 2014, pp. 437--440.
  [Online]. Available: \url{http://doi.acm.org/10.1145/2610384.2628055}
\BIBentrySTDinterwordspacing

\bibitem{DBLP:journals/corr/abs-1810-00314}
\BIBentryALTinterwordspacing
S.~Chakraborty, M.~Allamanis, and B.~Ray, ``Tree2tree neural translation model
  for learning source code changes,'' \emph{CoRR}, vol. abs/1810.00314, 2018.
  [Online]. Available: \url{http://arxiv.org/abs/1810.00314}
\BIBentrySTDinterwordspacing

\bibitem{He:2016:INM:3015812.3015835}
\BIBentryALTinterwordspacing
W.~He, Z.~He, H.~Wu, and H.~Wang, ``Improved neural machine translation with
  smt features,'' in \emph{Proc.\ of the 30th {AAAI} Conf. on Artificial
  Intelligence}, ser. AAAI'16.\hskip 1em plus 0.5em minus 0.4em\relax AAAI
  Press, 2016, pp. 151--157. [Online]. Available:
  \url{http://dl.acm.org/citation.cfm?id=3015812.3015835}
\BIBentrySTDinterwordspacing

\bibitem{AAAI1714451}
\BIBentryALTinterwordspacing
X.~Wang, Z.~Lu, Z.~Tu, H.~Li, D.~Xiong, and M.~Zhang, ``Neural machine
  translation advised by statistical machine translation,'' in \emph{AAAI
  Conference on Artificial Intelligence}, 2017. [Online]. Available:
  \url{https://aaai.org/ocs/index.php/AAAI/AAAI17/paper/view/14451}
\BIBentrySTDinterwordspacing

\bibitem{ArthurNN16}
P.~Arthur, G.~Neubig, and S.~Nakamura, ``Incorporating discrete translation
  lexicons into neural machine translation.'' in \emph{Proc.\ of Conference on
  Empirical Methods in Natural Language Processing}, ser. EMNLP'16, 2016, pp.
  1557--1567.

\bibitem{AAAI1714161}
\BIBentryALTinterwordspacing
Z.~Tu, Y.~Liu, L.~Shang, X.~Liu, and H.~Li, ``Neural machine translation with
  reconstruction,'' in \emph{AAAI Conference on Artificial Intelligence}, 2017.
  [Online]. Available:
  \url{https://aaai.org/ocs/index.php/AAAI/AAAI17/paper/view/14161}
\BIBentrySTDinterwordspacing

\bibitem{DBLP:journals/corr/LiuUFS16}
\BIBentryALTinterwordspacing
L.~Liu, M.~Utiyama, A.~M. Finch, and E.~Sumita, ``Neural machine translation
  with supervised attention,'' \emph{CoRR}, vol. abs/1609.04186, 2016.
  [Online]. Available: \url{http://arxiv.org/abs/1609.04186}
\BIBentrySTDinterwordspacing

\bibitem{DBLP:journals/corr/WuSCLNMKCGMKSJL16}
\BIBentryALTinterwordspacing
Y.~Wu, M.~Schuster, Z.~Chen, Q.~V. Le, M.~Norouzi, W.~Macherey, M.~Krikun,
  Y.~Cao, Q.~Gao, K.~Macherey, J.~Klingner, A.~Shah, M.~Johnson, X.~Liu,
  L.~Kaiser, S.~Gouws, Y.~Kato, T.~Kudo, H.~Kazawa, K.~Stevens, G.~Kurian,
  N.~Patil, W.~Wang, C.~Young, J.~Smith, J.~Riesa, A.~Rudnick, O.~Vinyals,
  G.~Corrado, M.~Hughes, and J.~Dean, ``Google's neural machine translation
  system: Bridging the gap between human and machine translation,''
  \emph{CoRR}, vol. abs/1609.08144, 2016. [Online]. Available:
  \url{http://arxiv.org/abs/1609.08144}
\BIBentrySTDinterwordspacing

\bibitem{Mizuno:2007:TEE:1287624.1287683}
\BIBentryALTinterwordspacing
O.~Mizuno and T.~Kikuno, ``Training on errors experiment to detect fault-prone
  software modules by spam filter,'' in \emph{Proc.\ of 6th Joint Meeting of
  the European Softw. Eng. Conf. and the {ACM} {SIGSOFT} Symp. on the Found. of
  Softw. Eng.}, ser. ESEC-FSE '07.\hskip 1em plus 0.5em minus 0.4em\relax New
  York, NY, USA: ACM, 2007, pp. 405--414. [Online]. Available:
  \url{http://doi.acm.org/10.1145/1287624.1287683}
\BIBentrySTDinterwordspacing

\bibitem{Hata:2008:EFF:1370750.1370772}
\BIBentryALTinterwordspacing
H.~Hata, O.~Mizuno, and T.~Kikuno, ``An extension of fault-prone filtering
  using precise training and a dynamic threshold,'' in \emph{Proc.\ of 5th
  Work. Conf. on Mining Softw. Repositories}, ser. MSR '08.\hskip 1em plus
  0.5em minus 0.4em\relax New York, NY, USA: ACM, 2008, pp. 89--98. [Online].
  Available: \url{http://doi.acm.org/10.1145/1370750.1370772}
\BIBentrySTDinterwordspacing

\bibitem{Tufano:2019:LMC:3339505.3339509}
\BIBentryALTinterwordspacing
M.~Tufano, J.~Pantiuchina, C.~Watson, G.~Bavota, and D.~Poshyvanyk, ``On
  learning meaningful code changes via neural machine translation,'' in
  \emph{Proc.\ of 41st Int. Conf. on Softw. Eng.}, ser. ICSE '19.\hskip 1em
  plus 0.5em minus 0.4em\relax Piscataway, NJ, USA: IEEE Press, 2019, pp.
  25--36. [Online]. Available: \url{https://doi.org/10.1109/ICSE.2019.00021}
\BIBentrySTDinterwordspacing

\bibitem{DBLP:journals/corr/abs-1901-01808}
\BIBentryALTinterwordspacing
Z.~Chen, S.~Kommrusch, M.~Tufano, L.~Pouchet, D.~Poshyvanyk, and M.~Monperrus,
  ``Sequencer: Sequence-to-sequence learning for end-to-end program repair,''
  \emph{CoRR}, vol. abs/1901.01808, 2019. [Online]. Available:
  \url{http://arxiv.org/abs/1901.01808}
\BIBentrySTDinterwordspacing

\bibitem{Saha:2019:HEM:3339505.3339508}
\BIBentryALTinterwordspacing
S.~Saha, R.~K. Saha, and M.~R. Prasad, ``Harnessing evolution for multi-hunk
  program repair,'' in \emph{Proc.\ of 41st Int. Conf. on Softw. Eng.}, ser.
  ICSE '19.\hskip 1em plus 0.5em minus 0.4em\relax Piscataway, NJ, USA: IEEE
  Press, 2019, pp. 13--24. [Online]. Available:
  \url{https://doi.org/10.1109/ICSE.2019.00020}
\BIBentrySTDinterwordspacing

\bibitem{Kim:2008:CSC:1399105.1399451}
\BIBentryALTinterwordspacing
S.~Kim, E.~J. Whitehead, Jr., and Y.~Zhang, ``Classifying software changes:
  Clean or buggy?'' \emph{{IEEE} Trans. Softw. Eng.}, vol.~34, pp. 181--196,
  March 2008. [Online]. Available:
  \url{http://portal.acm.org/citation.cfm?id=1399105.1399451}
\BIBentrySTDinterwordspacing

\bibitem{Hata:2012:BPB:2337223.2337247}
\BIBentryALTinterwordspacing
H.~Hata, O.~Mizuno, and T.~Kikuno, ``Bug prediction based on fine-grained
  module histories,'' in \emph{Proc.\ of 34th Int. Conf. on Softw. Eng.}, ser.
  ICSE '12.\hskip 1em plus 0.5em minus 0.4em\relax Piscataway, NJ, USA: IEEE
  Press, 2012, pp. 200--210. [Online]. Available:
  \url{http://dl.acm.org/citation.cfm?id=2337223.2337247}
\BIBentrySTDinterwordspacing

\bibitem{Ray:2016:NBC:2884781.2884848}
\BIBentryALTinterwordspacing
B.~Ray, V.~Hellendoorn, S.~Godhane, Z.~Tu, A.~Bacchelli, and P.~Devanbu, ``On
  the "naturalness" of buggy code,'' in \emph{Proc.\ of 38th Int. Conf. on
  Softw. Eng.}, ser. ICSE '16.\hskip 1em plus 0.5em minus 0.4em\relax New York,
  NY, USA: ACM, 2016, pp. 428--439. [Online]. Available:
  \url{http://doi.acm.org/10.1145/2884781.2884848}
\BIBentrySTDinterwordspacing

\bibitem{7588121}
D.~A. da~Costa, S.~McIntosh, W.~Shang, U.~Kulesza, R.~Coelho, and A.~E. Hassan,
  ``A framework for evaluating the results of the szz approach for identifying
  bug-introducing changes,'' \emph{{IEEE} Trans. Softw. Eng.}, vol.~43, no.~7,
  pp. 641--657, July 2017.

\bibitem{Tufano:2018:EIL:3238147.3240732}
\BIBentryALTinterwordspacing
M.~Tufano, C.~Watson, G.~Bavota, M.~Di~Penta, M.~White, and D.~Poshyvanyk, ``An
  empirical investigation into learning bug-fixing patches in the wild via
  neural machine translation,'' in \emph{Proc.\ of 33rd {IEEE}/{ACM} Int. Conf.
  on Automated Softw. Eng.}, ser. ASE 2018.\hskip 1em plus 0.5em minus
  0.4em\relax New York, NY, USA: ACM, 2018, pp. 832--837. [Online]. Available:
  \url{http://doi.acm.org/10.1145/3238147.3240732}
\BIBentrySTDinterwordspacing

\bibitem{2018arXiv181208693T}
M.~{Tufano}, C.~{Watson}, G.~{Bavota}, M.~{Di Penta}, M.~{White}, and
  D.~{Poshyvanyk}, ``{An Empirical Study on Learning Bug-Fixing Patches in the
  Wild via Neural Machine Translation},'' \emph{arXiv e-prints}, p.
  arXiv:1812.08693, Dec 2018.

\bibitem{2017arXiv170906182A}
M.~{Allamanis}, E.~T. {Barr}, P.~{Devanbu}, and C.~{Sutton}, ``{A Survey of
  Machine Learning for Big Code and Naturalness},'' \emph{ArXiv e-prints}, Sep.
  2017.

\bibitem{Allamanis:2013:MSC:2487085.2487127}
\BIBentryALTinterwordspacing
M.~Allamanis and C.~Sutton, ``Mining source code repositories at massive scale
  using language modeling,'' in \emph{Proc.\ of 10th Work. Conf. on Mining
  Softw. Repositories}, ser. MSR '13.\hskip 1em plus 0.5em minus 0.4em\relax
  Piscataway, NJ, USA: IEEE Press, 2013, pp. 207--216. [Online]. Available:
  \url{http://dl.acm.org/citation.cfm?id=2487085.2487127}
\BIBentrySTDinterwordspacing

\bibitem{Allamanis:2014:LNC:2635868.2635883}
\BIBentryALTinterwordspacing
M.~Allamanis, E.~T. Barr, C.~Bird, and C.~Sutton, ``Learning natural coding
  conventions,'' in \emph{Proc.\ of 22nd {ACM} {SIGSOFT} Int. Symp. on Found.
  of Softw. Eng.}, ser. FSE 2014.\hskip 1em plus 0.5em minus 0.4em\relax New
  York, NY, USA: ACM, 2014, pp. 281--293. [Online]. Available:
  \url{http://doi.acm.org/10.1145/2635868.2635883}
\BIBentrySTDinterwordspacing

\bibitem{Allamanis:2014:MIS:2635868.2635901}
\BIBentryALTinterwordspacing
M.~Allamanis and C.~Sutton, ``Mining idioms from source code,'' in \emph{Proc.\
  of 22nd {ACM} {SIGSOFT} Int. Symp. on Found. of Softw. Eng.}, ser. FSE
  2014.\hskip 1em plus 0.5em minus 0.4em\relax New York, NY, USA: ACM, 2014,
  pp. 472--483. [Online]. Available:
  \url{http://doi.acm.org/10.1145/2635868.2635901}
\BIBentrySTDinterwordspacing

\bibitem{Allamanis:2015:BMS:3045118.3045344}
\BIBentryALTinterwordspacing
M.~Allamanis, D.~Tarlow, A.~D. Gordon, and Y.~Wei, ``Bimodal modelling of
  source code and natural language,'' in \emph{Proc. of 32nd Int. Conf. on
  Machine Learning - Volume 37}, ser. ICML'15.\hskip 1em plus 0.5em minus
  0.4em\relax JMLR.org, 2015, pp. 2123--2132. [Online]. Available:
  \url{http://dl.acm.org/citation.cfm?id=3045118.3045344}
\BIBentrySTDinterwordspacing

\bibitem{Bielik:2016:PPM:3045390.3045699}
\BIBentryALTinterwordspacing
P.~Bielik, V.~Raychev, and M.~Vechev, ``Phog: Probabilistic model for code,''
  in \emph{Proc. of 33rd Int. Conf. on Machine Learning - Volume 48}, ser.
  ICML'16.\hskip 1em plus 0.5em minus 0.4em\relax JMLR.org, 2016, pp.
  2933--2942. [Online]. Available:
  \url{http://dl.acm.org/citation.cfm?id=3045390.3045699}
\BIBentrySTDinterwordspacing

\bibitem{Campbell:2014:SEJ:2597073.2597102}
\BIBentryALTinterwordspacing
J.~C. Campbell, A.~Hindle, and J.~N. Amaral, ``Syntax errors just aren't
  natural: Improving error reporting with language models,'' in \emph{Proc.\ of
  11th Work. Conf. on Mining Softw. Repositories}, ser. MSR 2014.\hskip 1em
  plus 0.5em minus 0.4em\relax New York, NY, USA: ACM, 2014, pp. 252--261.
  [Online]. Available: \url{http://doi.acm.org/10.1145/2597073.2597102}
\BIBentrySTDinterwordspacing

\bibitem{Cerulo:2015:IRI:2797418.2797556}
\BIBentryALTinterwordspacing
L.~Cerulo, M.~Di~Penta, A.~Bacchelli, M.~Ceccarelli, and G.~Canfora, ``Irish: A
  hidden markov model to detect coded information islands in free text,''
  \emph{Sci. Comput. Program.}, vol. 105, no.~C, pp. 26--43, Jul. 2015.
  [Online]. Available: \url{http://dx.doi.org/10.1016/j.scico.2014.11.017}
\BIBentrySTDinterwordspacing

\bibitem{Cummins:2017:SBP:3049832.3049843}
\BIBentryALTinterwordspacing
C.~Cummins, P.~Petoumenos, Z.~Wang, and H.~Leather, ``Synthesizing benchmarks
  for predictive modeling,'' in \emph{Proc. of 2017 Int. Symp. on Code
  Generation and Optimization}, ser. CGO '17.\hskip 1em plus 0.5em minus
  0.4em\relax Piscataway, NJ, USA: IEEE Press, 2017, pp. 86--99. [Online].
  Available: \url{http://dl.acm.org/citation.cfm?id=3049832.3049843}
\BIBentrySTDinterwordspacing

\bibitem{Gulwani:2014:NIP:2588555.2612177}
\BIBentryALTinterwordspacing
S.~Gulwani and M.~Marron, ``Nlyze: Interactive programming by natural language
  for spreadsheet data analysis and manipulation,'' in \emph{Proc. of 2014
  {ACM} {SIGMOD} Int. Conf. on Management of Data}, ser. SIGMOD '14.\hskip 1em
  plus 0.5em minus 0.4em\relax New York, NY, USA: ACM, 2014, pp. 803--814.
  [Online]. Available: \url{http://doi.acm.org/10.1145/2588555.2612177}
\BIBentrySTDinterwordspacing

\bibitem{Gvero:2015:SJE:2814270.2814295}
\BIBentryALTinterwordspacing
T.~Gvero and V.~Kuncak, ``Synthesizing java expressions from free-form
  queries,'' in \emph{Proc.\ of 2015 {ACM} {SIGPLAN} Int. Conf. on
  Object-Oriented Programming, Syst., Languages, and Appl.}, ser. OOPSLA
  2015.\hskip 1em plus 0.5em minus 0.4em\relax New York, NY, USA: ACM, 2015,
  pp. 416--432. [Online]. Available:
  \url{http://doi.acm.org/10.1145/2814270.2814295}
\BIBentrySTDinterwordspacing

\bibitem{Hata:2010:FMD:1743628.1743651}
\BIBentryALTinterwordspacing
H.~Hata, O.~Mizuno, and T.~Kikuno, ``Fault-prone module detection using
  large-scale text features based on spam filtering,'' \emph{Empirical Softw.
  Eng.}, vol.~15, pp. 147--165, April 2010. [Online]. Available:
  \url{http://dx.doi.org/10.1007/s10664-009-9117-9}
\BIBentrySTDinterwordspacing

\bibitem{Hellendoorn:2015:TLT:2820518.2820539}
\BIBentryALTinterwordspacing
V.~J. Hellendoorn, P.~T. Devanbu, and A.~Bacchelli, ``Will they like this?:
  Evaluating code contributions with language models,'' in \emph{Proc.\ of 12th
  Work. Conf. on Mining Softw. Repositories}, ser. MSR '15.\hskip 1em plus
  0.5em minus 0.4em\relax Piscataway, NJ, USA: IEEE Press, 2015, pp. 157--167.
  [Online]. Available: \url{http://dl.acm.org/citation.cfm?id=2820518.2820539}
\BIBentrySTDinterwordspacing

\bibitem{Hellendoorn:2017:DNN:3106237.3106290}
\BIBentryALTinterwordspacing
V.~J. Hellendoorn and P.~Devanbu, ``Are deep neural networks the best choice
  for modeling source code?'' in \emph{Proc.\ of 11th Joint Meeting on Found.
  of Softw. Eng.}, ser. ESEC/FSE 2017.\hskip 1em plus 0.5em minus 0.4em\relax
  New York, NY, USA: ACM, 2017, pp. 763--773. [Online]. Available:
  \url{http://doi.acm.org/10.1145/3106237.3106290}
\BIBentrySTDinterwordspacing

\bibitem{Hindle:2012:NS:2337223.2337322}
\BIBentryALTinterwordspacing
A.~Hindle, E.~T. Barr, Z.~Su, M.~Gabel, and P.~Devanbu, ``On the naturalness of
  software,'' in \emph{Proc.\ of 34th Int. Conf. on Softw. Eng.}, ser. ICSE
  '12.\hskip 1em plus 0.5em minus 0.4em\relax Piscataway, NJ, USA: IEEE Press,
  2012, pp. 837--847. [Online]. Available:
  \url{http://dl.acm.org/citation.cfm?id=2337223.2337322}
\BIBentrySTDinterwordspacing

\bibitem{Hsiao:2014:UWC:2660193.2660226}
\BIBentryALTinterwordspacing
C.-H. Hsiao, M.~Cafarella, and S.~Narayanasamy, ``Using web corpus statistics
  for program analysis,'' in \emph{Procl\ of 2014 {ACM} Int. Conf. on Object
  Oriented Programming Syst. Languages and Appl.}, ser. OOPSLA '14.\hskip 1em
  plus 0.5em minus 0.4em\relax New York, NY, USA: ACM, 2014, pp. 49--65.
  [Online]. Available: \url{http://doi.acm.org/10.1145/2660193.2660226}
\BIBentrySTDinterwordspacing

\bibitem{Karaivanov:2014:PST:2661136.2661148}
\BIBentryALTinterwordspacing
S.~Karaivanov, V.~Raychev, and M.~Vechev, ``Phrase-based statistical
  translation of programming languages,'' in \emph{Proc. of 2014 {ACM} Int.
  Symp. on New Ideas, New Paradigms, and Reflections on Programming and
  Software}, ser. Onward! 2014.\hskip 1em plus 0.5em minus 0.4em\relax New
  York, NY, USA: ACM, 2014, pp. 173--184. [Online]. Available:
  \url{http://doi.acm.org/10.1145/2661136.2661148}
\BIBentrySTDinterwordspacing

\bibitem{kushman-barzilay:2013:NAACL-HLT}
\BIBentryALTinterwordspacing
N.~Kushman and R.~Barzilay, ``Using semantic unification to generate regular
  expressions from natural language,'' in \emph{Proc. of 2013 Conf. of North
  American Chapter of Association for Computational Linguistics: Human Language
  Technologies}.\hskip 1em plus 0.5em minus 0.4em\relax Atlanta, Georgia:
  Association for Computational Linguistics, June 2013, pp. 826--836. [Online].
  Available: \url{http://www.aclweb.org/anthology/N13-1103}
\BIBentrySTDinterwordspacing

\bibitem{Maddison:2014:SGM:3044805.3044965}
\BIBentryALTinterwordspacing
C.~J. Maddison and D.~Tarlow, ``Structured generative models of natural source
  code,'' in \emph{Proc. of 31st Int. Conf. on Machine Learning - Volume 32},
  ser. ICML'14.\hskip 1em plus 0.5em minus 0.4em\relax JMLR.org, 2014, pp.
  II--649--II--657. [Online]. Available:
  \url{http://dl.acm.org/citation.cfm?id=3044805.3044965}
\BIBentrySTDinterwordspacing

\bibitem{Menon:2013:MLF:3042817.3042840}
\BIBentryALTinterwordspacing
A.~K. Menon, O.~Tamuz, S.~Gulwani, B.~Lampson, and A.~T. Kalai, ``A machine
  learning framework for programming by example,'' in \emph{Proc. of 30th Int.
  Conf. on Machine Learning - Volume 28}, ser. ICML'13.\hskip 1em plus 0.5em
  minus 0.4em\relax JMLR.org, 2013, pp. I--187--I--195. [Online]. Available:
  \url{http://dl.acm.org/citation.cfm?id=3042817.3042840}
\BIBentrySTDinterwordspacing

\bibitem{Nguyen:2013:SSL:2491411.2491458}
\BIBentryALTinterwordspacing
T.~T. Nguyen, A.~T. Nguyen, H.~A. Nguyen, and T.~N. Nguyen, ``A statistical
  semantic language model for source code,'' in \emph{Proc.\ of 9th Joint
  Meeting of the European Softw. Eng. Conf. and the {ACM} {SIGSOFT} Symp. on
  the Found. of Softw. Eng.}, ser. ESEC/FSE 2013.\hskip 1em plus 0.5em minus
  0.4em\relax New York, NY, USA: ACM, 2013, pp. 532--542. [Online]. Available:
  \url{http://doi.acm.org/10.1145/2491411.2491458}
\BIBentrySTDinterwordspacing

\bibitem{Nguyen:2015:DAM:2916135.2916174}
\BIBentryALTinterwordspacing
A.~T. Nguyen, T.~T. Nguyen, and T.~N. Nguyen, ``Divide-and-conquer approach for
  multi-phase statistical migration for source code (t),'' in \emph{Proc.\ of
  30th {IEEE}/{ACM} Int. Conf. on Automated Softw. Eng.}, ser. ASE '15.\hskip
  1em plus 0.5em minus 0.4em\relax Washington, DC, USA: IEEE Computer Society,
  2015, pp. 585--596. [Online]. Available:
  \url{https://doi.org/10.1109/ASE.2015.74}
\BIBentrySTDinterwordspacing

\bibitem{Nguyen:2015:GSL:2818754.2818858}
\BIBentryALTinterwordspacing
A.~T. Nguyen and T.~N. Nguyen, ``Graph-based statistical language model for
  code,'' in \emph{Proc.\ of 37th Int. Conf. on Softw. Eng. - Vol. 1}, ser.
  ICSE '15.\hskip 1em plus 0.5em minus 0.4em\relax Piscataway, NJ, USA: IEEE
  Press, 2015, pp. 858--868. [Online]. Available:
  \url{http://dl.acm.org/citation.cfm?id=2818754.2818858}
\BIBentrySTDinterwordspacing

\bibitem{Nguyen:2016:LAU:2884781.2884873}
\BIBentryALTinterwordspacing
T.~T. Nguyen, H.~V. Pham, P.~M. Vu, and T.~T. Nguyen, ``Learning api usages
  from bytecode: A statistical approach,'' in \emph{Proc.\ of 38th Int. Conf.
  on Softw. Eng.}, ser. ICSE '16.\hskip 1em plus 0.5em minus 0.4em\relax New
  York, NY, USA: ACM, 2016, pp. 416--427. [Online]. Available:
  \url{http://doi.acm.org/10.1145/2884781.2884873}
\BIBentrySTDinterwordspacing

\bibitem{Raychev:2014:CCS:2594291.2594321}
\BIBentryALTinterwordspacing
V.~Raychev, M.~Vechev, and E.~Yahav, ``Code completion with statistical
  language models,'' in \emph{Proc. of 35th {ACM} {SIGPLAN} Conf. on
  Programming Language Design and Implementation}, ser. PLDI '14.\hskip 1em
  plus 0.5em minus 0.4em\relax New York, NY, USA: ACM, 2014, pp. 419--428.
  [Online]. Available: \url{http://doi.acm.org/10.1145/2594291.2594321}
\BIBentrySTDinterwordspacing

\bibitem{Raychev:2016:LPN:2837614.2837671}
\BIBentryALTinterwordspacing
V.~Raychev, P.~Bielik, M.~Vechev, and A.~Krause, ``Learning programs from noisy
  data,'' in \emph{Proc. of 43rd Annual {ACM} {SIGPLAN-SIGACT} Symp. on
  Principles of Programming Languages}, ser. POPL '16.\hskip 1em plus 0.5em
  minus 0.4em\relax New York, NY, USA: ACM, 2016, pp. 761--774. [Online].
  Available: \url{http://doi.acm.org/10.1145/2837614.2837671}
\BIBentrySTDinterwordspacing

\bibitem{7081855}
A.~Sharma, Y.~Tian, and D.~Lo, ``Nirmal: Automatic identification of software
  relevant tweets leveraging language model,'' in \emph{2015 IEEE 22nd
  International Conference on Software Analysis, Evolution, and Reengineering
  (SANER)}, March 2015, pp. 449--458.

\bibitem{Tu:2014:LS:2635868.2635875}
\BIBentryALTinterwordspacing
Z.~Tu, Z.~Su, and P.~Devanbu, ``On the localness of software,'' in \emph{Proc.\
  of 22nd {ACM} {SIGSOFT} Int. Symp. on Found. of Softw. Eng.}, ser. FSE
  2014.\hskip 1em plus 0.5em minus 0.4em\relax New York, NY, USA: ACM, 2014,
  pp. 269--280. [Online]. Available:
  \url{http://doi.acm.org/10.1145/2635868.2635875}
\BIBentrySTDinterwordspacing

\bibitem{Vasilescu:2017:RCN:3106237.3106289}
\BIBentryALTinterwordspacing
B.~Vasilescu, C.~Casalnuovo, and P.~Devanbu, ``Recovering clear, natural
  identifiers from obfuscated js names,'' in \emph{Proc.\ of 11th Joint Meeting
  on Found. of Softw. Eng.}, ser. ESEC/FSE 2017.\hskip 1em plus 0.5em minus
  0.4em\relax New York, NY, USA: ACM, 2017, pp. 683--693. [Online]. Available:
  \url{http://doi.acm.org/10.1145/3106237.3106289}
\BIBentrySTDinterwordspacing

\bibitem{White:2015:TDL:2820518.2820559}
\BIBentryALTinterwordspacing
M.~White, C.~Vendome, M.~Linares-V\'{a}squez, and D.~Poshyvanyk, ``Toward deep
  learning software repositories,'' in \emph{Proc.\ of 12th Work. Conf. on
  Mining Softw. Repositories}, ser. MSR '15.\hskip 1em plus 0.5em minus
  0.4em\relax Piscataway, NJ, USA: IEEE Press, 2015, pp. 334--345. [Online].
  Available: \url{http://dl.acm.org/citation.cfm?id=2820518.2820559}
\BIBentrySTDinterwordspacing

\bibitem{Yadid:2016:ECP:2986012.2986021}
\BIBentryALTinterwordspacing
S.~Yadid and E.~Yahav, ``Extracting code from programming tutorial videos,'' in
  \emph{Proc. of 2016 {ACM} Int. Symp. on New Ideas, New Paradigms, and
  Reflections on Programming and Software}, ser. Onward! 2016.\hskip 1em plus
  0.5em minus 0.4em\relax New York, NY, USA: ACM, 2016, pp. 98--111. [Online].
  Available: \url{http://doi.acm.org/10.1145/2986012.2986021}
\BIBentrySTDinterwordspacing

\bibitem{Kim:2006:MBF:1181775.1181781}
\BIBentryALTinterwordspacing
S.~Kim, K.~Pan, and E.~E.~J. Whitehead, Jr., ``Memories of bug fixes,'' in
  \emph{Proc.\ of 14th {ACM} {SIGSOFT} Int. Symp. on Found. of Softw. Eng.},
  ser. SIGSOFT '06/FSE-14.\hskip 1em plus 0.5em minus 0.4em\relax New York, NY,
  USA: ACM, 2006, pp. 35--45. [Online]. Available:
  \url{http://doi.acm.org/10.1145/1181775.1181781}
\BIBentrySTDinterwordspacing

\bibitem{Nguyen:2010:RBF:1806799.1806847}
\BIBentryALTinterwordspacing
T.~T. Nguyen, H.~A. Nguyen, N.~H. Pham, J.~Al-Kofahi, and T.~N. Nguyen,
  ``Recurring bug fixes in object-oriented programs,'' in \emph{Proc.\ of 32nd
  Int. Conf. on Softw. Eng.}, ser. ICSE '10.\hskip 1em plus 0.5em minus
  0.4em\relax New York, NY, USA: ACM, 2010, pp. 315--324. [Online]. Available:
  \url{http://doi.acm.org/10.1145/1806799.1806847}
\BIBentrySTDinterwordspacing

\bibitem{Meng:2013:LLA:2486788.2486855}
\BIBentryALTinterwordspacing
N.~Meng, M.~Kim, and K.~S. McKinley, ``Lase: Locating and applying systematic
  edits by learning from examples,'' in \emph{Proc.\ of 35th Int. Conf. on
  Softw. Eng.}, ser. ICSE '13.\hskip 1em plus 0.5em minus 0.4em\relax
  Piscataway, NJ, USA: IEEE Press, 2013, pp. 502--511. [Online]. Available:
  \url{http://dl.acm.org/citation.cfm?id=2486788.2486855}
\BIBentrySTDinterwordspacing

\bibitem{8094441}
R.~Yue, N.~Meng, and Q.~Wang, ``A characterization study of repeated bug
  fixes,'' in \emph{Proc.\ of 33rd {IEEE} Int. Conf. on Softw. Maintenance and
  Evolution}, Sept 2017, pp. 422--432.

\end{thebibliography}
%





\end{document}